\documentclass[%
 aip,
 amsmath,amssymb,
 reprint,%
]{revtex4-1}

\usepackage{color}
\usepackage{graphicx}
\usepackage{dcolumn}
\usepackage{bm}
\usepackage{soul}

\usepackage[utf8]{inputenc}
\usepackage[T1]{fontenc}
\usepackage{mathptmx}
\usepackage{etoolbox}
\usepackage{xcolor}

\newcommand{\dd}{\mathrm{d}}

\makeatletter
\def\@email#1#2{%
 \endgroup
 \patchcmd{\titleblock@produce}
  {\frontmatter@RRAPformat}
  {\frontmatter@RRAPformat{\produce@RRAP{*#1\href{mailto:#2}{#2}}}\frontmatter@RRAPformat}
  {}{}
}%
\makeatother
\begin{document}

\preprint{AIP/123-QED}

\title[Plasma propulsion simulation using particles]{Plasma propulsion simulation using particles}
\author{F. Taccogna}
 \affiliation{ Institute for Plasma Science and Technology (ISTP), CNR, Bari, Italy}%
 \email{francesco.taccogna@cnr.it}
\author{F. Cichocki}
\affiliation{ Institute for Plasma Science and Technology (ISTP), CNR, Bari, Italy}%

\author{D. Eremin}
\affiliation{Institute of Theoretical Electrical Engineering, Ruhr University Bochum, Universitätsstraße 150, 44801 Bochum, Germany}%

\author{G. Fubiani}
\affiliation{LAPLACE, Universit\'e de Toulouse, CNRS, INPT, UPS, Toulouse, France}%
 
\author{L. Garrigues}
\affiliation{LAPLACE, Universit\'e de Toulouse, CNRS, INPT, UPS, Toulouse, France}%

\date{\today}

\begin{abstract}
This perspective paper deals with an overview of particle-in-cell / Monte Carlo collision models applied to different plasma-propulsion configurations and scenarios, from electrostatic ($\bm{E} \times \bm{B}$ and pulsed arc) devices to electromagnetic (RF inductive, helicon, electron cyclotron resonance) thrusters, with an emphasis on plasma plumes and their interaction with the satellite. The most important elements related to the modeling of plasma-wall interaction are also presented.
Finally, the paper reports new progress in the particle-in-cell computational methodology, in particular regarding accelerating computational techniques for multi-dimensional simulations and plasma chemistry Monte Carlo modules for molecular and alternative propellants.
\end{abstract}

\maketitle

%

\section{\label{sec: introduction} Introduction}

Electric propulsion (EP) has been developed since the early 60s, and its use onboard satellites, orbiting platforms, and interplanetary probes has increased significantly in the last two decades \cite{Ahedo2011,Mazouffre2016,lev19,Jorns23}. The need for a detailed understanding of the working physics and a precise estimation of the performance to enable cheaper innovative designs has spurred the development of a large number of simulation codes, with particle-based codes \cite{BirdsallLangdon,Bird} playing a significant role.

Plasma thrusters can be classified in terms of the gas ionization process, the basic conversion mechanism for the kinetic energy gained by the ions \cite{andr03}, the main acceleration mechanism of the plasma \cite{Ahedo2011}, or the modeling needs. In this perspective paper on EP numerical simulations using particles, we shall consider the last classification. From this point of view, electric thrusters (with the exception of electro-thermal thrusters) fall within the electrostatic (ES) and electromagnetic (EM) categories. Thrusters belonging to the former can be modeled by retaining only Poisson's equation, while the second category requires including the full set (or a subset) of Maxwell's equations.

ES thrusters are based on the acceleration of positive charges under the action of a DC electric field, with a cathode emitting electrons being used to neutralize the positive charge flow, thus preventing the space vehicle charging \cite{Mazouffre2016}. Most of ES thrusters feature a cylindrical geometry and an electrical power input varying from hundreds of W to tens of kW. In ES thruster concepts such as the gridded ion engines (GIE) \cite{Holst20} and $\bm{E} \times \bm{B}$ based devices [see, e.g., Hall thrusters - HT \cite{Zhurin1999,Goeb08,Boeuf17}, cylindrical HT \cite{Rait01}, highly efficient multistage plasma thruster - HEMPT \cite{Korn03}, cusped-field thruster - CFT \cite{MacDonald_2011}, and co-axial magneto-isolated anode - CAMILA \cite{LEV2016}], positive ions are generated under the ionization of a propellant gas (usually Xe) by high energetic electrons, and are accelerated by an imposed potential difference between two electrodes. In the case of GIEs, the electrodes are represented by a system of polarized electrostatic grids negatively biased with respect to the plasma potential and featuring multi-apertures. In $\bm{E} \times \bm{B}$ thrusters, like the HT, the potential difference is imposed between an inner anode and an external cathode, and the DC electric field that accelerates the ions is sustained inside the discharge channel under the action of an applied magnetic field.
The growth of the micro and nano-satellites market has fostered the development of alternative device concepts to deal with the low electrical power available (less than 100 W). Colloid thrusters (CT) - also called electrospray thrusters - enter in the category of ES engines, although positive charges are now liquid droplets produced by an electrospray process \cite{HUANG2021}. Finally, in pulsed arc thrusters (PAT), the plasma expands after being formed from the ablation (in pulsed plasma thruster - PPT) or the evaporation (in vacuum arc thruster - VAT) of a solid propellant grain \cite{KEIDAR2018}. PATs can be considered ES thrusters because the magnetic field induced by the plasma currents is generally negligible compared to either the externally applied one (in VATs) or the one induced by the external circuit current (in PPTs).

Differently, electromagnetic (EM) thrusters ionize and accelerate the propellant under the combined action of magnetic and time-dependent electric fields, and, in some cases, electromagnetic waves \cite{Ahedo2011, Mazouffre2016, O'Reilly2021}. Therefore, the necessity of solving for the full (reduced in some cases) set of Maxwell's equation is associated with two different reasons: the type of electron heating for plasma generation and the type of plasma acceleration. Regarding the former, power absorption can be due to RF induction heating (as in inductive plasma thrusters IPT \cite{Takao2010, Henrich2013, Dubois2018, Polzin2020, takao2023}) or wave heating (as in RF Helicon wave \cite{Takahashi2019} HPT, microwave electron cyclotron resonance \cite{Cannat2015, Takao2014, Sanchez2021} ECRT and ion cyclotron resonances VASIMR \cite{Olsen2015} thrusters). This sub-category of EM thrusters is electrodeless, thus having the advantage of limiting the plasma-wall interaction and mitigating the problem of their lifetime. Regarding plasma acceleration, since EM thrusters are quasi-neutral (except inside the Debye sheaths at the walls), this is related to a $\bm{j} \times \bm{B}$ term (hence the naming Lorentz Force Accelerators LFA), and $\bm{B}$ can present, in some cases, a significantly large self-induced component (and hence the need of including, at least, Ampère's law). This is the case of the magneto-plasma-dynamic thrusters (MPDT), where the magnetic induction field can be either applied (AF \cite{Krulle98}) or self-induced (SF \cite{Li2019}).

Today, the design and development of electric thrusters are still semi-empirical and require long and expensive life tests. There is a need to better understand key plasma processes occurring in the complex partially magnetized plasmas, such as the electron heating, electron and ion transport, the plasma-wall interaction, or the ion-induced erosion, and to address the question of alternative propellants.
The experimental characterization of electric thrusters in ground-test facilities has some intrinsic criticalities: i) the difficulty of having reliable diagnostics on devices that often do not allow direct and non-invasive access; ii) the influence of the ground-test facilities on thruster performance (operations of the thruster are very sensitive to the chamber background pressure); iii) the reproducibility of the natural thruster working conditions, such as the typical space environment conditions or flow (for atmospheric breathing EP concept). In this regard,  numerical simulations play a crucial role. Moreover, it is also of paramount importance to better understand the discharge configuration and the emitted plasma plume's interaction with the spacecraft for realistic geometries, which clearly requires the development of 3D numerical tools. If these were made available to the industry, the efficiency of existing products could be increased, and new design breakthroughs would be enabled.

In recent years, numerical simulations have increasingly benefited the basic understanding and engineering optimization of electric thrusters.  This is due to several concurrent contributions: (i) the evolution of computer hardware that has allowed the representation of multi-dimensional geometries and multi-scale phenomena avoiding numerical tricks (e.g., artificial vacuum permittivity, mass ratio, size scaling, slower speed of light), (ii) the implementation of sophisticated new algorithms and numerical diagnostic tools and (iii) the availability of new collisional and surface interaction data.

In particular, the kinetic and non-local description is indispensable to capture the intrinsic nature of the different EP concepts, since low-pressure partially magnetized plasmas feature a large number of inherently kinetic, non-equilibrium, and non-local phenomena. Among the different kinetic approaches \cite{Colonna22} available, the particle representation allows a simple and intuitive implementation to account for a detailed description of plasma-gas and plasma-wall interactions, although it suffers from the inevitable but controllable limitation due to inherent discrete particle noise \cite{Turner06, Riva17, Tavassoli21, Vass22} and from its computationally intensive nature. This perspective article critically discusses the different particle-based methodologies used to describe the plasma (electron and ion species) and coupled gas phases in both the discharge and plume regions of the most common EP configurations.

The paper is organized as follows. Sec.\,\ref{sec: particle-based models} introduces the most commonly used particle-based models for the plasma and gas species. Then, the different Particle-in-Cell schemes for the plasma discharge simulations are presented in Sec.\,\ref{sec: pic models thrusters}. Models available for plasma plumes simulation, satellite interaction and plume-electromagnetic compatibility are presented in Sec.\,\ref{sec: plumes and SC interaction}, while future simulation challenges are addressed in Sec.\,\ref{sec: future challenges}. Finally conclusions are drawn in Sec.\,\ref{sec: conclusions}.

Finally, we assembled a very comprehensive list of references that should be credited for the original work.

\section{\label{sec: particle-based models} Particle-in-Cell and Monte Carlo simulations}
The Particle-in-Cell (PIC) technique  \cite{Hockney_1988,BirdsallLangdon,Tskhakaya07,Brieda19,Donko21,Colonna22} is a Lagrangian/Eulerian (or particle-mesh) method, applicable to low temperature and low pressure discharges like those of electric thrusters that are characterized by a weakly coupled and low collisional plasma, exhibiting non-equilibrium behavior and non-local properties.

The PIC approach dates back to the late 50s when pioneering simulations of Buneman \cite{Buneman1959} and Dawson \cite{Dawson1962} have been implemented to study some basic properties of collisionless plasmas. A few years later, the first stochastic Monte Carlo simulations of the charged particle transport have been applied to drift tubes \cite{Skullerud1968,Lin1977} and gas discharges \cite{Sakai1977,Pitchford1982,Boeuf1982}, while aerodynamicists developed a particle-particle simulation method for the neutral species collisions \cite{Bird}, known as Direct Simulation Monte Carlo (DSMC). Subsequently, the PIC approach started to be coupled with Monte Carlo methods for the simulation of collisional processes \cite{Birdsall1991,Nanbu00} also in plasma discharges, with the first applications to plasma thrusters appearing in 90s thanks to the contributions of Arakawa's \cite{Komurasaki1995} and Martinez-Sanchez's \cite{Lenz1993,Fife1998} research groups and Adam-Héron \cite{Garrigues00}.

\begin{figure}[!ht]
    \centering \includegraphics[width=0.47\textwidth]{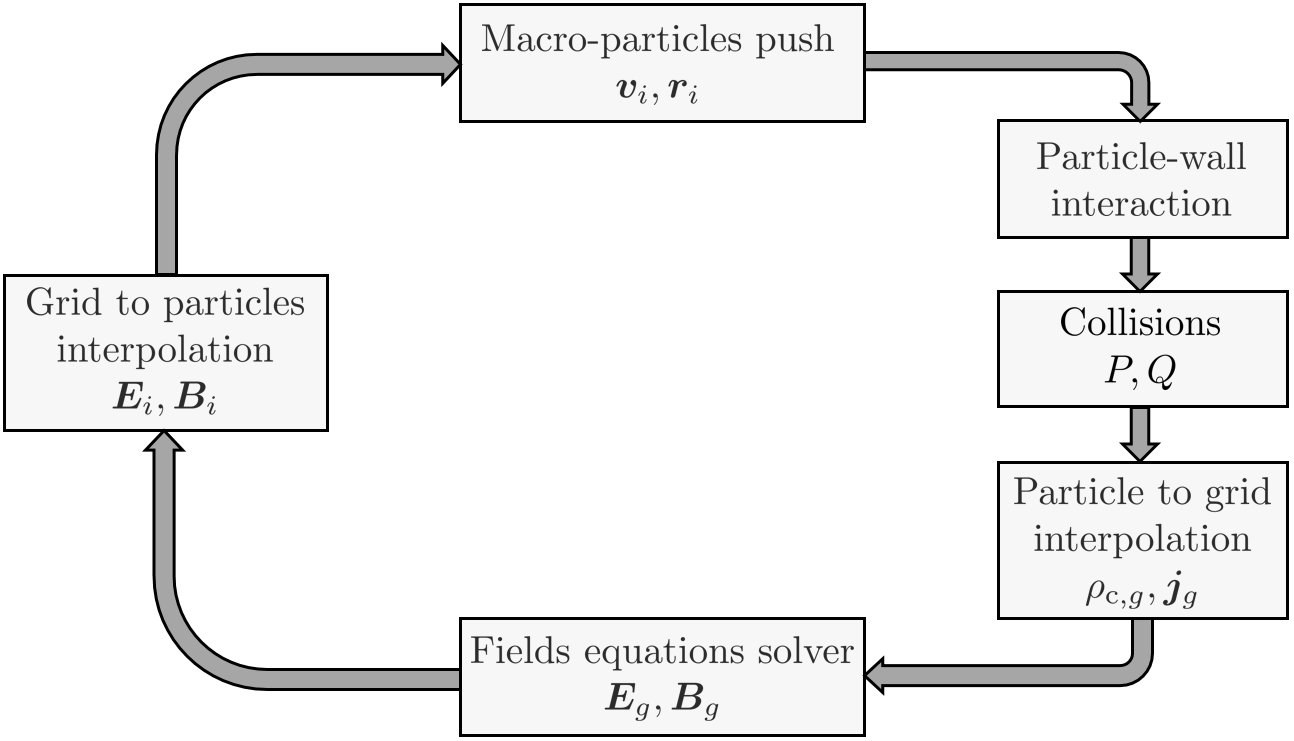}
    \caption{Scheme of a typical PIC-Monte Carlo loop}
    \label{fig: PIC Montecarlo cycle}
\end{figure}

In PIC / Monte Carlo models, the distribution function $f$ of both the plasma (ions and electrons) and the neutral gas species is approximated by an ensemble of $N$ macro-particles (or super-particles) as:
\begin{equation}
\label{eq: mp}
f(\bm{r},\bm{v}) = \sum_{p=1}^N w_p S(\bm{r}-\bm{r}_p) \delta (\bm{v}-\bm{v}_p)
\end{equation}
where $w_p$ is the statistical weight of the $p^\mathrm{th}$ macro-particle (many models considering a uniform weight for all particles of a given species), $\bm{r}_p,\bm{v}_p$ are its position and velocity vector, $\delta$ is the Dirac delta function and $S$ is the macro-particle cloud shape function determining how the macro-particle weight is distributed in space and satisfying the integral relation $\iiint S(\bm{r}') \dd^3 r' = 1$. As illustrated in Fig.\,\ref{fig: PIC Montecarlo cycle}, the PIC/Monte Carlo method consists in solving the corresponding species (electron, ion and atom/molecule) Boltzmann's equations featuring the following steps:

\begin{enumerate}
    \item The trajectories of the macro-particles between collisional events are obtained by solving the Newton's equation: 
    \begin{align}
    \begin{cases}
    \label{eq: EoM}
    m_p \dfrac{\mathrm{d}\bm{v}_p}{\mathrm{d}t} & = q_p (\bm{E}_p + \bm{v}_p \times \bm{B}_p), \\
    \dfrac{\mathrm{d}\bm{r}_p}{\mathrm{d}t} & = \bm{v}_p ,
    \end{cases}
    \end{align}
    where $q_p$ and $m_p$ are the elementary particle charge and mass, respectively, and $\bm{E}_p,\bm{B}_p$ are the electric and magnetic fields at the macro-particle location, obtained as:
    \begin{align}
    \begin{cases}
    \bm{E}_p = & \sum_{g} \bm{E}_g W(\bm{r}_g-\bm{r}_p), \\
    \bm{B}_p = & \sum_{g} \bm{B}_g W(\bm{r}_g-\bm{r}_p),
    \end{cases}
    \end{align}
    being $W(\bm{r}) = \iiint { S\left(\bm{r}'\right) \Pi \left(\bm{r} - \bm{r}'\right) \dd^3 r'}$ the interpolation function or assignment function shape  (being $\Pi$ the rectangular function with width equal to the mesh spacing \cite{Hockney_1988}), and $\bm{E}_g$, $\bm{B}_g$ the known electric and magnetic fields at the grid point $g$. The trajectories of the macro-particles are obtained through the integration of the discretized form of Eq.\,\eqref{eq: EoM} over a time step $\Delta t$ (leap frog methods being the most diffuse \cite{Hockney_1988, Birdsall1991}).

    \item The macro-particles may reach the physical walls or computational domain boundaries, where various processes (absorption, reflection, emission of additional particles) may take place as defined by the plasma-surface and gas-surface models. Here again different probabilistic Monte Carlo models can be used depending on the process implemented (see Sec.\,\ref{subsec: plasma wall interaction}). This step may also include any particle injection algorithm, which is often necessary to maintain the plasma.

    \item The collisions of the particles are handled through the Nanbu (no-time counter) stochastic Monte Carlo rules \cite{Nanbu00}:
    \begin{equation}
    \label{eq: MC}
    f(t+\Delta t) = (1-P)f(t) +PQ
    \end{equation}
    with the following probabilistic interpretation: a particle with velocity $\bm{v}_p$ will not collide with probability $(1-P)$, and it will collide with probability $P$, according to the collision laws described by $Q$. Different Monte Carlo schemes have been set up for electron-molecule, ion-molecule, molecule-molecule and Coulomb collisions \cite{Nanbu00}. These can be roughly classified in two main categories: (i) Monte Carlo Collisions (MCC) methods which consider a fluid collisional background for the ``projectile'' macro-particles, and (ii) DSMC methods, in which collisions between macro-particle pairs are actually considered. 
    
    \item The charge and current densities of macro-particles as well as the electric/magnetic fields are computed on a discrete grid. The latter quantities are obtained from Maxwell's equations with a dedicated solver that takes into account the space charge density $\rho_\mathrm{c}$ and current density $\bm{j}$ generated by the ensemble of charged particles at each grid point $g$,
    \begin{align}
    \begin{cases}
    \rho_{\mathrm{c},g} = & \dfrac{1}{V_g} \sum_{p} q_p w_p W(\bm{r}_g-\bm{r}_p), \\
    \bm{j}_g = & \dfrac{1}{V_g} \sum_{p} q_p w_p \bm{v}_p W(\bm{r}_g-\bm{r}_p),
    \end{cases}
    \label{eq: deposition}
    \end{align}
    with $V_g$ representing the cell volume associated to the grid point, as well as the effects of any external power source that appear as boundary conditions of Maxwell's equations.
\end{enumerate}
The PIC/Monte Carlo method is therefore characterized by a combination of different algorithms characterized by their own accuracy order: the pusher (time-integrator) to solve the mesh-free equation of motions, the field solver for the mesh-based Maxwell's equations, the interpolation schemes to couple mesh-free quantities with the mesh-based ones, and the stochastic Monte Carlo rules for volumetric binary collisions and surface processes. All parts are important and we have to deal with their numerical error to reduce the full error of the PIC method. The physical constraints, such as the conservation of mass, momentum and energy are important to the physical experiments and should be conserved by the underlying numerical schemes.

The standard PIC scheme described above, has then to be adapted to the different problem, e.g. an electrostatic (ES) or electromagnetic (EM) one. Furthermore, the different numerical ideas to solve the time-dependencies, e.g. explicit or implicit schemes (see Sec.\,\ref{subsec: reducing cost}), has a strong influence on the numerical stability of PIC codes \cite{Vass22}. For example, in EM codes, an explicit solver with integration time step $\Delta t$ has to satisfy $\omega_\mathrm{p,e} \Delta t <$ 2, where $\omega_\mathrm{p,e}$ is the electron plasma frequency, or the Courant-Friedlich-Levy condition $c \leq \Delta r/\Delta t$, where $c$ is the phase wave velocity and $\Delta r$ the mesh spacing. Furthermore the grid solver has to satisfy the electron Debye length $\Delta r \leq \zeta \lambda_\mathrm{D,e}$, where $\lambda_\mathrm{D,e}$ is the Debye length and $\zeta$ is a constant of order 1. Finally, the collision probabilities should be sufficiently small, in order to minimize the effect of ``missed'' collisions (i.e. multiple collisions during a time step). It is a good practice to keep the collision probability $P$ of a given species below 5-10\% to limit the error to less than 1\% \cite{Donko21,vahe95}.

The particle representation enable to efficiently handle the multidimensionality associated with non-equilibirum discharges and considerable flexibility to model advanced physics (such as internal energy excitation, chemical reactions, surface interactions, etc.).

\section{\label{sec: pic models thrusters} PIC models of thruster discharges}

\subsection{\label{subsec: electrostatic PIC} Electrostatic PIC}
In electrostatic thrusters, the effects of self-induced fields are generally considered negligible compared to the applied external fields. The magnetic induction field $\bm{B}(\bm{r},t)$ is prescribed in both space and time, and is not coupled with the plasma currents. Therefore, the PIC model for macro-particles is uniquely coupled with Poisson's equation for the electrostatic potential $\phi$:
\begin{equation}
  \label{eq: poisson equation}
  \epsilon_0 \nabla^2 \phi =  e \left( n_\mathrm{e} - \sum \limits_{s} { Z_s n_s} \right),
\end{equation}
where $\epsilon_0$ is the vacuum dielectric constant, $n_\mathrm{e}$ is the electron number density, $e$ is the elementary charge, and $Z_s, n_s$ are respectively the charge number and number density of the considered ion species. Therefore, at each time step, from the knowledge of the deposited number densities of electron and ion species, a Poisson's solver is employed to obtain the electric potential, and hence the self-consistent electrostatic field $\bm{E}_\mathrm{ES} = - \nabla {\phi}$. This is then interpolated from PIC mesh nodes to the particles position to advance them in the following time step, as shown in Fig.\,\ref{fig: PIC Montecarlo cycle}.

Eq.\,\eqref{eq: poisson equation} is discretized with methods of varying complexity (in some cases accounting for adaptive mesh refinement AMR), from Finite Differences to Finite Volumes methods. The resulting linear system is solved with highly efficient sparse solvers, employing either direct (e.g. PARDISO \cite{sche04}, MUMPS \cite{ames01}) or iterative methods (e.g, PETSc \cite{bala19}), with the latter being more convenient in expensive 3D geometries. The assumed boundary conditions for $\phi$ are of either Dirichlet or Neumann type: the former are applied to boundary surfaces whose potential is known a priori (e.g. at the electrodes), while the latter are used at either open boundaries (refer to Sec.\,\ref{subsec: PIC models for plumes} for more details) or at the dielectric surfaces. In this latter case, a common approximation is to neglect the electric field inside the material and impose a capacitor-like condition based on the local surface charge density $\sigma_\mathrm{c}$:
\begin{equation}
\label{eq: dielectric boundary condition}
\nabla \phi \cdot \bm{1}_\mathrm{n}=-\sigma_\mathrm{c}/\epsilon_0,
\end{equation}
where $\bm{1}_\mathrm{n}$ represents the unit vector pointing toward the plasma. However, recent works \cite{Taccogna2022} have also considered the effect of the relative permittivity $\epsilon_\mathrm{r}$ of any considered dielectric material by solving the generalized version of Poisson's Eq.\,\eqref{eq: poisson equation},
\begin{equation}
  \label{eq: poisson equation1}
  \nabla \cdot (\epsilon_\mathrm{r} \nabla \phi) =  - \frac{\rho_\mathrm{c}}{\epsilon_0},
\end{equation}
in a extended domain that includes the dielectrics in order to correctly compute the electric field discontinuities at the plasma-dielectric interface \cite{Nagel14}.

In the vast majority of ES PIC simulations existing in literature, an easy-to-implement explicit scheme is used (refer to Sec.\,\,\ref{subsec: reducing cost} for alternative implicit schemes), so that the strong constraints presented in Sec.\, \ref{sec: particle-based models} exist on the grid spacing and time step resolution to avoid numerical instabilities, with the only difference that the CFL condition is now based on the fastest species, the electrons, i.e. $v_\mathrm{e} \leq \Delta r/\Delta t$. The resulting computational cost can be huge, depending on the considered thrusters. For PAT thrusters, the plasma density peaks to $10^{22}$ $\rm {m^{-3}}$, corresponding to a minimum Debye length $\lambda_\mathrm{D,e} \approx \uppercase{O}(10^{-7})$m, and hence to a time step $\Delta t \approx \uppercase{O}(10^{-13})$s (from CFL condition, having assumed an electron temperature of $\approx$ 10 eV). Since meaningful simulations should cover at least a few $\mu$s, approximately $10^8$ time steps are to be completed. In the other ES thrusters ($\bm{E} \times \bm{B}$, GITs or CTs), the plasma density level is lower, with a typical maximum value around $10^{18}$ ${\rm m^{-3}}$, yielding $\lambda_\mathrm{D,e} \approx \uppercase{O}(10^{-5})$m and $\Delta t \approx \uppercase{O}(10^{-11})$ s, respectively. When neutrals dynamics is simulated, the steady state is reached after fractions of ms so that a total of around $10^7$ time steps is generally required. For the above reasons, and especially in 3D simulations, it is therefore necessary to use supercomputers and High Performance Computing (HPC) techniques (see Sec.\,\ref{subsec: hpc techniques}). Nevertheless, numerical tricks (e.g. enlarged vacuum permittivity and/or reduction of ion mass or thruster size scaling) are still used in many simulations to reduce the computational time (see Refs.\,\onlinecite{Coche14, Zhao_2014, Yongjie_2016, Tacco18, Kahnfeld_2018, Matyash_2019} for $\bm{E} \times \bm{B}$ thrusters, and Refs.\,\onlinecite{Lusk18, Yang2020, LIU_2019} for PATs).

In the following paragraphs, some peculiarities of the different ES thruster types are further discussed.

\paragraph{PAT thrusters} PAT thrusters are generally simulated by assuming a magnetic induction field $\bm{B}=\bm{B}(\bm{r},t)$ that is not coupled with the plasma (hence the validity of the electrostatic approximation). In VATs \cite{Lusk18,Yang2020}, this is non-uniform in space but assumed to be constant in time, while in PPTs, it also varies with time \cite{LIU_2019}. In particular, it is obtained as a function of the current flowing in the external circuit between the capacitor plates, and of the instantaneous position of the accelerated plasma beam packet, with the use of Biot–Savart formulas \cite{LIU_2019}. Therefore, even for PPTs, no self-consistent solution of Ampère's law, accounting for the plasma currents, is generally considered, at least in PIC models. 
The neutral atoms transport, including ablation or evaporation, collisions and surface-interaction (see Sec.\, \ref{subsec: plasma wall interaction} for more information) is then described with a kinetic approach \cite{Lusk18, LIU_2019}. The use of complex mixtures of solid propellant materials leads to a large level of uncertainty in terms of the collisional input data (see Sec.\,\ref{subsec: alternative propellants}), if available at all in literature. Moreover, the production of multiply charged ions from different species induced by the high-voltage [for example chlorine and carbon ions for polytetrafluoroethylene (PTFE) used as propellant for PPT \cite{LIU_2019}] at very high densities leads to an obvious increase of the computational time, making it necessary to reduce the simulation cost with the usual numerical tricks cited above \cite{Yang2020}.

\paragraph{Electro-spray thrusters}
In these thrusters, the magnetic induction field is generally absent $\bm{B} = \bm{0}$. Narayanan \textit{et al.} \cite{Nara17} have modeled the transport of charged particles emitted from a source at one end of the computational domain expanding through an aperture positioned downstream and polarized at a negative voltage. The size is of the aperture is two orders of magnitude larger than the emission source. They have implemented an adaptive mesh refinement technique to reduce the computational time. Zao \textit{et al.} \cite{Zhao19}, on the other hand, have simulated the droplet acceleration using a particle-particle method (where the droplets are modeled with spherical particles) and the electrostatic force between them using Coulomb's interaction. This method avoids the use of mesh and constraints associated and Poisson's equation resolution issues. Fundamental studies have finally also addressed the question of the formation of droplets through molecular dynamic (MD) techniques \cite{Wang13, enom22}. These studies can provide more precise injection conditions for particle-based models of electro-spray thrusters, which currently consider a thermal injection of ions/liquid droplets.

\paragraph{Gridded ion engines}
\begin{figure*}[!ht]
	\centering
	\includegraphics[width=1\textwidth]{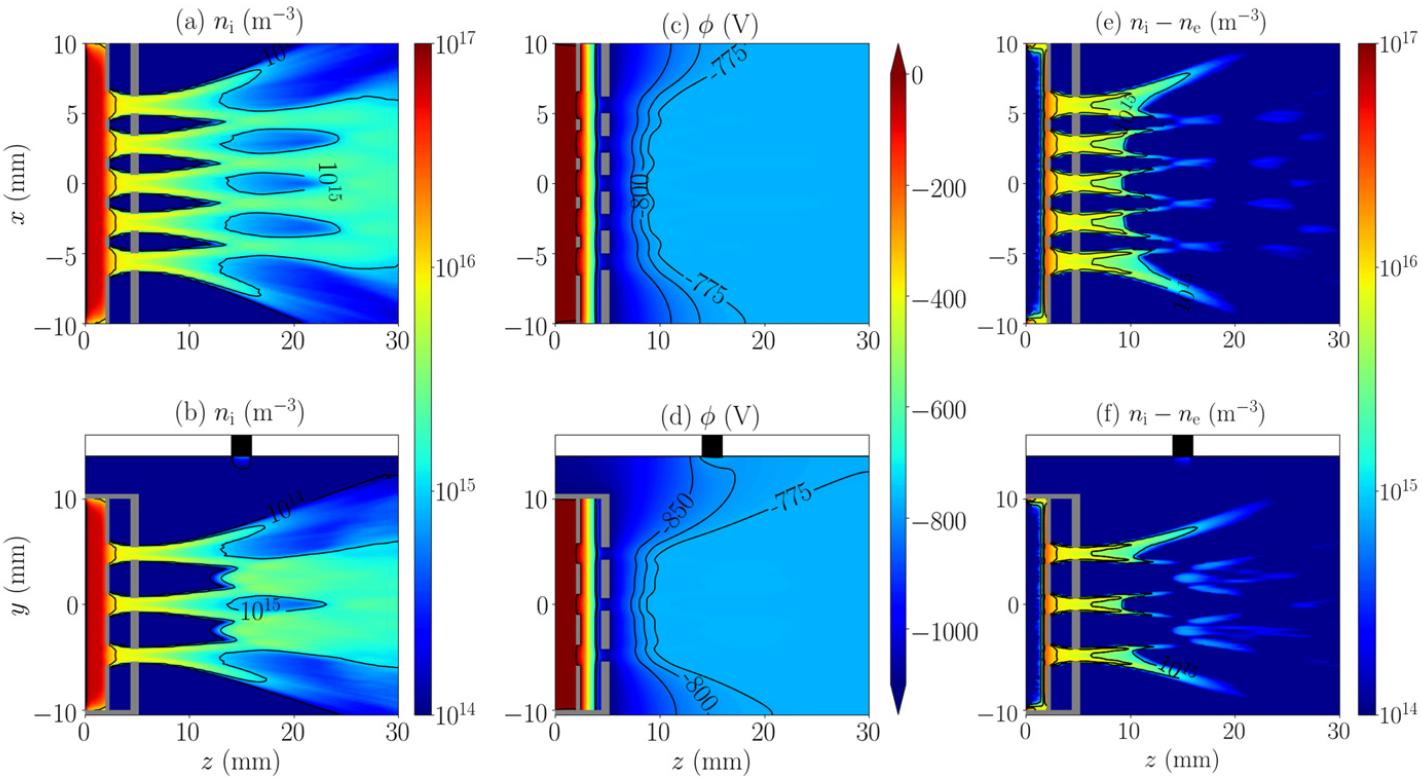}
        \caption{Three dimensional simulation of beamlets extraction, acceleration and coalescence in a GIE with multiple apertures, obtained with a hybrid model. Subplots (a,c,e) and (b,d,f) refer to different thruster cross sections. Reproduced with permission from Perales \textit{et al.}\cite{pera21}, Plasma Sources Science and Technology \textbf{30}, 105023 (2021). Copyright 2021 IOP Publishing Ltd.}
	\label{fig: ion grid optics simulation}
\end{figure*}

In GIE simulations, the magnetic induction field is either absent or assumed to be constant in time and prescribed in space, $\bm{B}=\bm{B}(\bm{r})$. Although simulations of very similar negative ion sources for neutral beam injection (NBI) for nuclear fusion reactors are abundant in literature \cite{Taccogna10,Taccogna11,Taccogna13,tacc17,Fubiani17,Garrigues23}, simulations covering the interior of gridded ion engines are not very diffuse in the plasma propulsion community, although a few examples exist where, the ion generation region (sometimes called ``driver'' region) is not self-consistently simulated \cite{yama18}, but accounted for as an effective ``volumetric injection region''. More commonly, GIE simulations assume given upstream plasma conditions (in terms of plasma density and electron temperature) and focus exclusively on the ion beam extraction and acceleration process. This is indeed fundamental for both the design of the extraction grids and for the estimation of their ion-induced sputtering levels, and hence the ion thruster lifetime. Moreover, most times, a hybrid approach is used \cite{jian15,pera21} (refer to Sec.\, \,\ref{sec: plumes and SC interaction} A for a more detailed description), in which the electron density is given by a simplified Boltzmann relation with a fixed electron temperature (sometimes two electron populations are considered for the separated upstream and downstream plasma electrons), alleviating the constrains related to the electrons dynamics. Three-dimensional codes, including commercial ones, are then capable of simulating realistic conditions, including a large number of grid apertures. They are now sufficiently mature to be used for the optimization and design of the accelerating grids (thickness, spacing, and grid voltages) minimizing the ion impacts on the grids (see Ref.\,\onlinecite{Holst20} for a more complete overview of the available tools).
Fig.\,\ref{fig: ion grid optics simulation} finally shows the results for the ion extraction, acceleration and beamlets coalescence of a multi-apertures GIE, obtained with a hybrid PIC model, featuring fluid electrons.

\begin{figure}[!ht]
	\centering
	\includegraphics[scale=2.7]{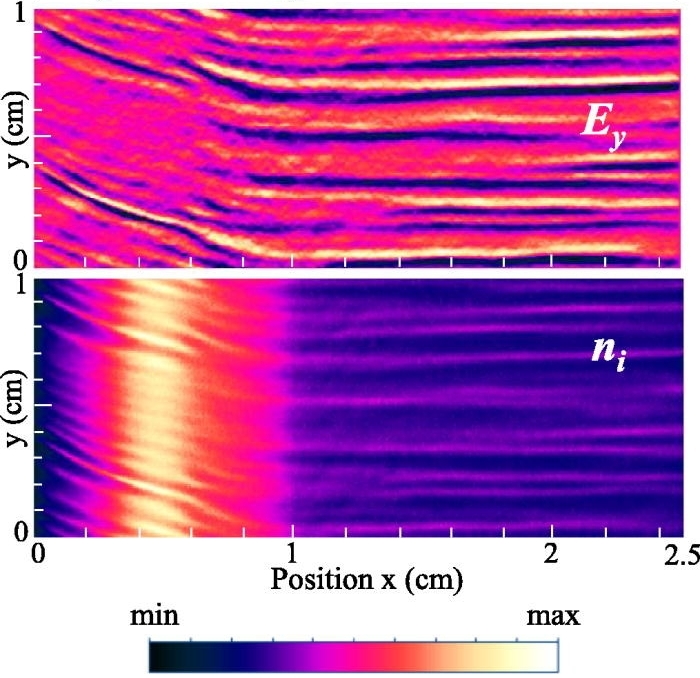}
        \caption{Axial \emph{x} - azimuthal \emph{y} distributions of the azimuthal electric field $E_\mathrm{y}$ and of the ion density $n_\mathrm{i}$ at a given time step. The minimum and maximum values for $E_\mathrm{y}$ are $-5\times{{10}^{4}}$ and $5\times{{10}^{4}}$ $\text{V/m}$, respectively, and for $n_\mathrm{i}$ are 0 and $5\times{{10}^{17}}$ $\text{m}^{-3}$. Adapted with permission from Boeuf \textit{et al.}. \cite{Boeuf18}, Phys. Plasmas \textbf{25}, 061204 (2018). Copyright 2018 AIP Publishing LLC.}
	\label{figII.A.1}
\end{figure}

\paragraph{$\bm{E}\times \bm{B}$ thrusters}
In these thrusters, the magnetic induction field is prescribed and constant in time $\bm{B}=\bm{B}(\bm{r})$. Although hybrid PIC models with fluid electrons assume a particle description for the neutrals (as in Refs.\,\onlinecite{,Hagelaar02,Panelli21,pera22}), in fully kinetic simulations, the propellant gas generally follows a fluid description. It is modeled as a non-uniform background with given density and temperature conditions \cite{Coche14, Charoy_2019, Charoy_2021, Petronio2023}, which are considered fixed in time, in order to reduce the computational time. Scaling of the thruster size, and prescribing the neutral density profile \cite{Croes_2017, Tacco18, Taccogna_2019, Tavant_2018} or the source term profile \cite{Boeuf18} are commonly used techniques and permit to study specific processes and to assess their influence on the thruster operations. Even if these studies have shown limited conclusions, they offer opportunities to make parametric analyses minimizing the computational cost, before gradually increasing the complexity including the coupling between different physical phenomena. 
A typical example is the study of the electron cyclotron drift instability (ECDI) and its role on the so-called anomalous electron transport. This kinetic instability \cite{Lafl18} takes place in the azimuthal $\bm{E} \times \bm{B}$ direction of a HT and is due to the different velocities between magnetized electrons (rotating azimuthally under the action of the axial electric and radial magnetic fields) and ions, that, on the other hand remain unmagnetized and are accelerated almost purely in the axial direction. 
One-dimensional azimuthal \cite{Boeuf14, Lafl2016, Janh18a, Katz2018, Asad19}, two-dimensional axial-azimuthal \cite{Boeuf14, Charoy_2019, Char20, Charoy_2021, Coche14, Katz2018, Lafl18, Taccogna_2019}, and two-dimensional radial-azimuthal \cite{Janh18b, Croes_2017, Taccogna_2019, Villafana_2021, Tavant_2018, Petr21, Seng20} models have been developed. Simulation results show the existence of a coherent structure in the azimuthal direction with resonances at discrete values characterized by a wavelength in the order of the millimeter and a frequency in the MHz range (as illustrated in Fig.~\ref{figII.A.1}). 
The saturation of the instability is visible in ES PIC simulations with the deformation and broadening of the electron and ion velocity distribution functions. The origin of the transition to the non-linear regime is still in debate (see Ref.\,\onlinecite{TacGar_2019} for more details). 2D radial-azimuthal simulations have also revealed the existence of a longer wavelength instability between the dielectric walls along the magnetic field line, called Modified Two-Stream Instability (MTSI). These 2D radial-azimuthal models have also included the secondary electron emission (SEE) induced by plasma electrons colliding with the walls to study the mutual coupling between the ECDI and SEE \cite{Taccogna_2019, Tavant_2018}. Lastly, three-dimensional ES PIC simulations of a HT have been performed under simplified plasma conditions \cite{mine17, Tacco18,  Kaganovich20, VilPhD2021, Villa2022}. Preliminary analyses have shown the existence of the ECDI but whose fluctuations amplitude is typically one order of magnitude smaller than in 2D simulations, close to the estimations built with Thomson scattering measurements \cite{Tsik10}.

\begin{figure*}[!ht]
	\centering
	\includegraphics[scale=1.0]{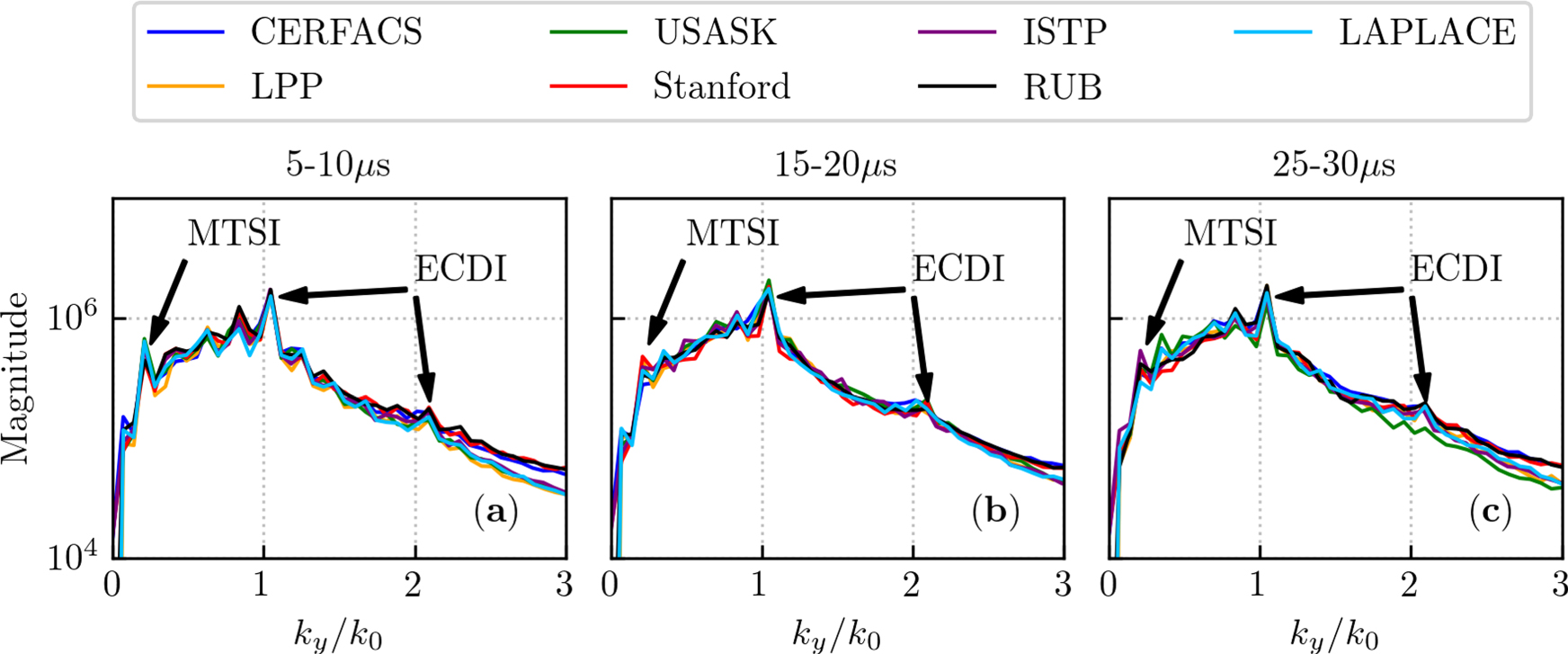}
        \caption{1D azimuthal \uppercase{FFT} of the azimuthal electric field $\uppercase{E}_\text{y}$, averaged over all radial positions and over three temporal intervals obtained by seven ES PIC models implemented by different research groups. MTSI and ECDI resonances are indicated by arrows. Reproduced with permission from Villafana \textit{et al.}\cite{Villafana_2021},  Plasma Sources Science and Technology \textbf{30}, 075002 (2021). Copyright 2021 IOP Publishing Ltd.}
	\label{figII.A.2}
\end{figure*}

Finally, the community has been interested in recent years in the verification aspect of the numerical techniques used for $\bm{E} \times \bm{B}$ thrusters, through the LANDMARK project \cite{LAND}. This aspect is essential to demonstrate the accuracy of the simulations. These benchmarks, reproduced by seven research groups worldwide, also allowed a comparative evaluation to examine the efficiency of the implementation of PIC algorithms (type of parallelization, Poisson's solvers, explicit vs implicit algorithms, CPU compared to GPU architectures, etc. \cite{Charoy_2019,Villafana_2021}). Fig.\,\ref{figII.A.2} illustrates a comparison of the two-dimensional radial-azimuthal simulation results implemented by seven different ES PIC models showing their capability to capture the time evolution of the MTSI and ECDI instabilities with a minimal difference.

\subsection{\label{subsec: electromagnetic PIC} Electromagnetic PIC}

EM thruster discharges are often characterized by a relatively large plasma density ($n_\mathrm{e} \geq 10^{18}$ m$^{-3}$) and low electron temperature ($T_\mathrm{e} \leq$ 10 eV), which leads to small values of the plasma period $\Delta T_\mathrm{p,e}= 1/\omega_\mathrm{p,e}$ and Debye length $\lambda_\mathrm{D,e}$. If one tried to simulate such discharges with a conventional momentum-conserving explicit PIC method, the corresponding limitations on the cell size and time step (note that in the case of EM simulations, one also has to respect the very restrictive CFL condition, which requires resolving the time a light wave traverses a grid cell) would result in too large computational times. Therefore, such discharges are often modeled either by the two-fluid \cite{Takao2006,Takahashi2009,Magarotto2020,Magarotto2022} or by hybrid \cite{Dubois2018,Sanchez2021} (fluid for electrons and particle for ions, see Sec.\, \ref{sec: plumes and SC interaction} A) descriptions. However, owing to the fact that thrusters are operated at low pressures and Coulomb collisions are relatively weak, kinetic and non-local effects become necessary to account for. Furthermore, the finite electron inertia effects, typically omitted in fluid or hybrid models, might play a crucial role in capturing the propagation of EM waves in the discharge plasmas and other relevant effects correctly. The PIC method satisfies all these criteria; nevertheless, only recently and for miniaturized configurations, some fully kinetic PIC descriptions have been attempted.

One can use the Helmholtz theorem to decompose the electric field into the potential (rotation-free) and solenoidal (divergence-free) parts \cite{Zangwill2013}, which can be represented through the scalar and vector potentials as $-\nabla\phi$ and $-\partial \bm{A}/\partial t$, respectively, with the vector potential $\bm{A}$ satisfying the Coulomb gauge condition, $\nabla\cdot\bm{A}=0$. If the driving frequency $\omega$ is small enough so that the corresponding EM wavelength is much bigger than the system size, then the electrostatic approximation applies and the potential part of the electric field, which is obtained from Poisson's equation, dominates. That is why this part is often referred to as the ``electrostatic field'' $\bm{E}_\mathrm{ES}$, even when the ES approximation cannot be used and the fully EM treatment is demanded. In this case, one must be careful, as the solenoidal component, which is referred to as the ``inductive electric field'' $\bm{E}_\mathrm{EM}$, also contains a potential part besides the inductive part. The latter is needed to enforce the finite signal propagation speed, and thus causality since the electrostatic solution describes an instantaneous change in the entire space \cite{Jackson1999}. Such a decomposition can be convenient if there is a certain symmetry expected, which permits to employ the fully EM description only for selected directions, whereas the coupled field-plasma dynamics in the other directions can be treated with the ES approximation (e.g., Ref.\,\onlinecite{Takekida2006}; see also ``Inductively coupled thrusters'' section below). Another case warranting the use of such a technique could be a low-power plasma discharge with $\omega_\mathrm{p,e}\ll\omega$, when the pumping EM wave is hardly affected by the plasma, and the only effect is a weak damping on plasma electrons (see, e.g., Refs.\onlinecite{Takao2014,Yamashita2019,Fu2022}). In this case, the solenoidal $\bm{E}_\mathrm{EM}$ component will describe the fast EM pumping wave pre-computed in a vacuum, with the amplitude adjusted according to the power absorbed in the plasma. The potential $\bm{E}_\mathrm{ES}$ part will describe predominantly quasi-stationary ES ambipolar fields governing slow plasma diffusion and diamagnetic currents if a magnetic field is present. However, in order to neglect the plasma response at the microwave frequencies, the plasma should not be dense, i.e., $\omega_{pe}\ll \omega$ should hold, which is rarely fulfilled. This could lead to the omission of many important effects \cite{Geller1996}.

In a general situation, where large time-modulated electric fields can arise due to space charge effects (such as in the plasma sheaths of a capacitively-coupled plasma CCP discharge\cite{Eremin2022} or in the plasma bulk of a helicon discharge for the $m=0$ mode \cite{Chen2015}), dropping the potential part of the displacement current (see Eq.\,\eqref{EMPIC_eq3}) and solving for $\bm{E}_\mathrm{ES}$ and $\bm{E}_\mathrm{EM}$ separately is not justified, because they are coupled. In contrast, in many situations, the magnetic field can be split into the static $\bm{B}_\mathrm{ES}$ (note that in this case ``ES'' is a misnomer, which is frequently used in the literature though) and the time-modulated $\bm{B}_\mathrm{EM}$ parts. The latter is often neglected in magnetized discharges due to the dominance of the static magnetic field. 

The equation of motion of the $p^\mathrm{th}$ macro-particle, Eq.\,\eqref{eq: EoM}, can be written as:
\begin{equation}
m_p \frac{\mathrm{d}\bm{v}_p}{\mathrm{d}t} = q_p \left[\bm{E}_\mathrm{ES}+\bm{E}_\mathrm{EM}+\bm{v}_p \times (\bm{B}_\mathrm{ES}+\bm{B}_\mathrm{EM})\right] 
\label{EMPIC_eq1},
\end{equation}
coupled with Faraday's and Ampère's equations:
\begin{equation}
\frac{\partial \bm{B}_\mathrm{EM}}{\partial t} = - \nabla \times \bm{E}_\mathrm{EM} 
\label{EMPIC_eq2}
\end{equation}
\begin{equation}
\varepsilon_0 \frac{\partial \bm{E}_\mathrm{EM} } {\partial t} = \frac{1}{\mu_0} \nabla \times \bm{B}_\mathrm{EM} - (\bm{j}_\mathrm{ext} + \bm{j}), 
\label{EMPIC_eq3}
\end{equation}
where $\mu_0$ is the vacuum permeability, while $\bm{j}_\mathrm{ext}$ is the current density inside the antenna/coil generating the RF electric field/microwaves in plasma, and $\bm{j}$ represents the plasma current, which is generated predominantly by electrons since the ion motion is negligible on the driving RF/microwave timescale. Note that to have a general applicability, Eq.\,\eqref{EMPIC_eq3} must also include the $\varepsilon_0 \partial \bm{E}_\mathrm{ES}/\partial t$ term on the left-hand-side (needed to ensure the charge continuity in Ampère's law) so that the decomposition of the electric field into the potential and solenoidal parts is redundant because the electric field calculated from the Maxwell equations contains both parts in a natural way. If a PIC algorithm is charge-conserving, i.e., it ensures a correct coupling between the current and the charge density in the sense of the continuity equation for the chosen shape functions, and an algorithm ensures that the induced $\bm{B}$ field is purely solenoidal, then one obtains the correct EM fields from Faraday's and Ampère's laws. Should this not be the case, the Gauss laws for the electric and magnetic fields should be explicitly enforced.

When the feedback of the plasma on the field is important, it is necessary to solve for the full $\bm{E}$ and $\bm{B}$ fields from the complete set of Maxwell's equations without making any approximations \cite{Koh93,Gopinath95,Pavarin2011,Jaafarian2018,Porto2021,Porto2022,Porto2023}. Direct time integration of the coupled field-plasma system using the simplest leapfrog-based explicit EM PIC scheme is typically intractable due to the CFL restriction $\Delta r > \min(c\Delta t,\lambda_\mathrm{D,e})$. This is why most works reported in the literature rely on certain PIC model reductions to eliminate the stiffness in the numerical particle orbit and the field integration algorithms leading to the CFL condition.
A popular workaround is to substitute the classical finite-difference time-domain (FDTD) algorithm \cite{Yee66,Verboncoeur1995} by solving for the EM fields $\bm{E}_\mathrm{EM}$ and $\bm{B}_\mathrm{EM}$ induced by the RF antenna in the frequency domain using the Helmholtz's equation \cite{Takao2011,takao2023}
\begin{equation}
\big (\nabla^2+\mu_0\varepsilon_0\omega^2 \big ) \tilde{\bm{E}}_\mathrm{EM} = i\omega\mu_0(\tilde{\bm{j}}_\mathrm{ext}+\tilde{\bm{j}}) \label{EMPIC_eq4},
\end{equation}
where $\tilde{\bm{E}}_\mathrm{EM}$ is the complex electric field, $i$ is the imaginary unit, and $\tilde{\bm{j}}$ is the complex plasma current density obtained, at grid point $g$ of the PIC mesh (see Eq.\,\eqref{eq: deposition}), as
\begin{equation}
\tilde{\bm{j}}_{g} = \dfrac{\sum_{p}^{} q_p w_p \bm{v}_p}{V_g} e^{i\Delta\psi}
\label{EMPIC_eq5}
\end{equation}
where $\Delta \psi$ is the phase difference between the RF field and the plasma current. Since $\bm{j}_\mathrm{ext}$ is usually non-zero only at the boundary or outside of the computational domain, the information on it is incorporated in the boundary conditions imposed on $\tilde{\bm{E}}_\mathrm{EM}$.
The electromagnetic field $\bm{B}_\mathrm{EM}$ is then directly calculated from Faraday’s law, Eq.\,\eqref{EMPIC_eq2}. Frequently, only the response at the driving frequency is taken into account, and the corresponding amplitude is updated every time interval of the order of the RF period\cite{Henrich2013}, which is much bigger than the time step. The underlying assumption that the plasma response at the stationary state is dominated by the linear contribution at the fundamental driving frequency harmonic thus excludes certain non-linear or transient time effects. Non-linear effects, such as the generation of higher harmonics, parametric decay instabilities, non-linear wave-particle interaction including particle trapping effects, non-linear skin-effect (see, e.g., Refs.\onlinecite{Erokhin1973,smolyakov2000,ostrikov2003,Kline2003,Chen2006,Sydorenko2006}) can be important but require evolving the EM fields in the time domain. Despite the computational cost of the conventional explicit EM PIC, it could still be possible to employ it in some cases\cite{Gopinath95}, when the corresponding thruster is small enough or by artificially reducing it using scaling techniques (in the latter case, one has to be cautious though, as the physics might not be entirely equivalent to that of the original setup \cite{Yuan2020,Petronio2023}). The continuous development of advanced HPC techniques (see \ref{subsec: hpc techniques} for more details) and hardware makes simulations with the explicit PIC method feasible for larger devices. A conducive feature of the EM PIC method is that the leapfrog algorithm evolving the EM fields does not involve matrix inversions and can be very efficiently parallelized so that the fields can be evolved with a much smaller time step than the particles to avoid the CFL restriction caused by the field evolution algorithm. 
Alternatively, one can use field time integration algorithms that are not sensitive to the time step. Here, two different options are possible. On the one hand, one can eliminate some of the fast time scales from the model by making certain simplifying assumptions. An appropriate example would be the PIC algorithms based on the electric field estimates resulting from the quasi-neutrality ansatz rather than Poisson's equation, used for ECR\cite{Lampe1998} and ICP\cite{Sydorenko2005} discharge modeling, which removes plasma oscillations. On the other hand, one can use numerical algorithms that can tolerate large time steps by design. The semi-Lagrangian Constrained Interpolation Profile (CIP) algorithm \cite{Porto2021} employed in Refs.\onlinecite{Porto2022,Porto2023} is one of the possible approaches that could be mentioned in this context. Being explicit, it is relatively simple to implement and does not incur a heavy computational burden. Additionally, it has good numerical dissipation and dispersion properties. Another option is to use implicit PIC methods (see Sec. \ref{subsec: implicit PIC} for more details). Such methods seem to be surprisingly rarely employed in EM PIC modeling of plasma propulsion devices. However, the recently developed energy-conserving PIC (ECPIC), or the semi-implicit PIC (ECSIM) methods demonstrate very attractive properties when applied to EM modeling of plasmas featuring a large ratio between the system size and the Debye length \cite{Mattei2017,Eremin2022}. 

An adequate self-consistent description of the plasma and the EM field dynamics is vital for obtaining accurate power absorption profiles. The latter are intimately related to the profiles of plasma density, electrostatic potential, electron temperature, and ionization rate, which affect other properties of the discharge of immediate relevance to the propulsion applications, such as the electron confinement and transport, plume dynamics, thrust, and many others. It is also important to be able to predict the specific energy acquired per electron from the electric field in the context of electron energization, i.e., production of electrons with energy above the ionization threshold \cite{anders_2014,Eremin2023a,Eremin2023b}.
However, the claims that in the end, it is the global power coupling that determines the plasma properties \cite{Godyak2013} and that the plasma power absorption, density, and temperature profiles are very similar in various kinds of plasma discharges \cite{godyak2020}, appear to be an oversimplification \cite{Takahashi2020}.  
Each type of discharge is different and deserves a separate study.
Although it is true that one can get insights into some of the thruster-related phenomena without having to model the EM field and power absorption dynamics and using some simplifying assumptions or even prescribed profiles, a consistent and comprehensive understanding followed by quantitative estimates and optimizations can result only when the EM field propagation is appropriately treated.

The following thruster types have been modeled most using EM PIC codes:

\paragraph{Inductively coupled thrusters \label{ICP}}
Thrusters based on inductively coupled plasma discharges are among the oldest EP technologies\cite{Loeb2005} remaining in active use nowadays\cite{Holst20}. They typically feature a planar coil with an RF current, which creates a time-dependent magnetic field with both radial and axial components, inducing an azimuthal electric field sustaining the plasma discharge. This ``inductive heating'' mode (also known as the H heating mode) occurs at relatively large powers, when the skin depth is smaller than the discharge size. At low powers, the discharge is operated in the ``capacitive heating'' mode (also known as the E heating mode), which is powered by the radial and axial capacitive electric fields produced by the coil. In the absence of a stationary magnetic field, the thrust is generated by ions extracted from the system by electrostatic grids, e.g., as in radio-frequency ion thrusters (RIT).

The azimuthal electric field drives a current, which, when combined with the magnetic field, generates second harmonics of the current radial and azimuthal components via the Lorentz force\cite{Godyak1999,Godyak1999a,ostrikov2003}. The latter, in turn, generates the second harmonic of the azimuthal magnetic field. Such a non-linear coupling process can be significant at lower RF frequencies and can generate ever higher harmonics of the magnetic field\cite{smolyakov2000}. The induced magnetic field can affect electron orbits, which is important when kinetic and non-local phenomena play out, leading to the anomalous skin effect and the ponderomotive force\cite{Sydorenko2005,Smolyakov2007}. Under typical conditions of this type of thruster plasmas, only the second harmonics matter, which justifies the common assumption that it is sufficient to describe only the azimuthal component of the electric field electromagnetically, using Eqs. (\ref{EMPIC_eq4}) and (\ref{EMPIC_eq5}), whereas the radial and the axial electric field components can be described electrostatically, using Poisson's equation\cite{Takao2010,Takao2011,Takao2012,Takao2015,Henrich2013,Henrich2017}. Eq.\,\eqref{EMPIC_eq4} should be supplemented by the boundary conditions, which can be obtained from the observation that in this case $\tilde{\bm{E}}_\mathrm{EM}(\bm{r}) = -i \omega \tilde{\bm{A}}(\bm{r})$, where the vector potential can be calculated from the Biot-Savart's law (assuming that the electromagnetic radiation wavelength is much larger than the distance from the coil to the boundary of the computational domain)
\begin{equation}
\tilde{\bm{A}}(\bm{r}) = \frac{\mu_0}{4\pi}\int d^3r^\prime \frac{(\tilde{\bm{j}}_\mathrm{ext}(\bm{r}^\prime)+\tilde{\bm{j}}(\bm{r}^\prime))}{|\bm{r}-\bm{r}^\prime|}.
\label{EMPIC_eq6}
\end{equation}
Although it is expensive to calculate this integral in general, it has to be calculated only at the boundary. As mentioned before, calculation of the electric field $\tilde{\bm{E}}_\mathrm{EM}(\bm{r})$ in the Fourier space avoids the CFL problems and reduces computational costs because it needs to be calculated only once per RF period\cite{Henrich2013}.

Regarding PIC simulations of the propulsion-related ICP plasma discharges, Ref.\,\onlinecite{Takao2010} compared experimental, fluid, and kinetic results for a discharge in argon at pressure 370–770 mTorr, driven at 450 MHz and RF powers below 3.5 W, and found discrepancies between the 2D fluid and the PIC results, which can be linked to the non-Maxwellian EEDF observed in the PIC simulations. In Ref.\,\onlinecite{Takao2011}, a similar discharge in Xe at 4.2 mTorr was investigated with the same 2D PIC approach for the range of driving frequencies 5–500 MHz. It was concluded that lower RF frequencies lead to higher plasma densities and more uniform plasma profiles. Refs.\,\onlinecite{Henrich2013,Henrich2017} described a 3D PIC model, which was massively parallelized on CPUs and could simulate complex geometries. Selected plasma properties obtained from the simulations of a $\mu$N-RIT 1.0 were shown. Ref.\,\onlinecite{Takao2012} demonstrated the importance of accounting for the capacitive coupling. Ref.\,\onlinecite{Takao2015} explored the ion momentum loss to the lateral wall and showed that it is enhanced due to the depletion of the neutral gas.

\paragraph{MW-driven ECR plasma thruster \label{MWECR}}
The microwave (MW) heating mechanism for electric thruster is often associated with the electron cyclotron resonance (ECR) configuration having a magnetic field to increase the energy transfer efficiency of the wave to the plasma. The reason is that at low pressures, the plasma resonance in an unmagnetized plasma is too weak. In contrast, a magnetic field leads to the possibility of having a resonance at the cyclotron frequency or its harmonics and can confine electrons (possibly, in combination with the electrostatic potential) so that they pass through that resonance many times, gaining more energy from the electric field. Due to the thermal motion of the electrons, the resonance condition becomes a Doppler shift, and for a population of electrons having a distribution in velocity space, the resonance location becomes Doppler-broadened\cite{Williamson1992}.
The energy absorption via ECR heating occurs orthogonally to the magnetic field, which results in a strongly anisotropic electron energy distribution function (EEDF). In addition, close to the resonance, the electric field amplitude becomes large, and in a warm plasma, Langmuir and/or Bernstein waves can be excited \cite{Geller1996}. These waves carry the energy from the resonance and can deposit it somewhere else in the plasma through their own damping mechanisms. Such a mode conversion process can be linear when the secondary waves are excited at the same frequency as that of the driving wave, or it can be non-linear when different harmonics are generated, which can result in turbulence. The thrust in such systems can be generated either by employing a system of grids accelerating the ions or by a magnetic nozzle, where ions are accelerated by the ambipolar field sustained by the plasma. Evidently, most of the mentioned phenomena should be treated kinetically and nonlocally.

\begin{figure*}
    \centering
    \includegraphics[scale=1.5]{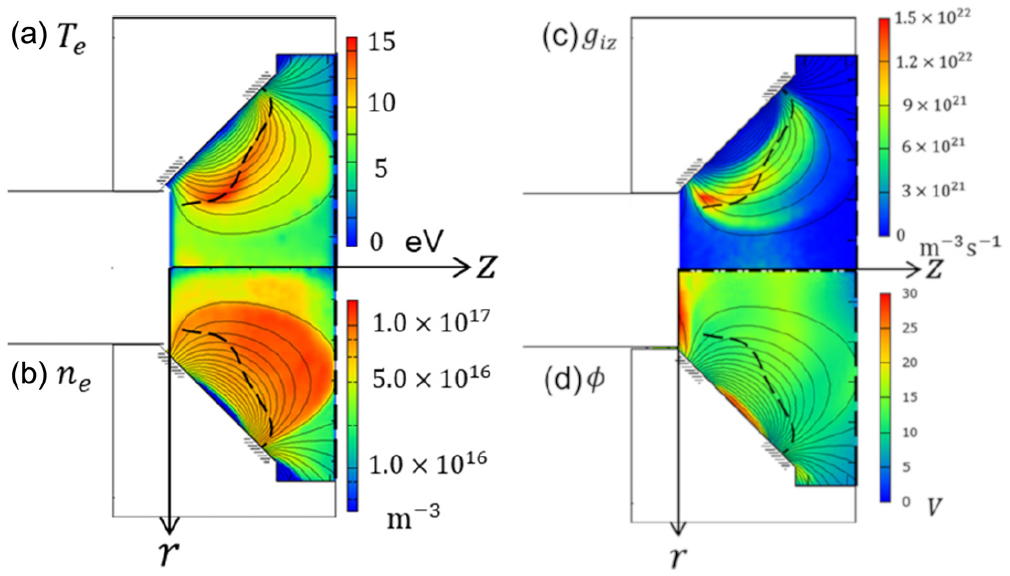}
    \caption{The $(r,z)$ profiles of electron temperature, electron density, ionization rate, and electrostatic potential in the $\mu$10 ECR thruster. Reproduced with permission from Yamashita \textit{et al.}\cite{Yamashita2019}, Physics of Plasmas \textbf{26}, 073510 (2019), Copyright 2019 AIP Publishing.}
    \label{fig: ECR thruster properties}
\end{figure*}

The first works, which modeled MW-ECR ion sources by using one-dimensional \cite{Koh93} and three-dimensional \cite{Gopinath95} EM PIC simulations, date back to the 90s. They considered the propagation of the pumping wave and the power absorption profiles in 1D\cite{Koh93} and 3D\cite{Gopinath95}, respectively. The latter was possible due to the use of the HPC techniques available at that time. The corresponding PIC approaches described the self-consistent evolution of the plasma and EM fields in the time domain. The coupled integration of the particle and field equations in the ime domain in the propulsion-related PIC simulations of ECR discharges was recently employed in Refs.\,\onlinecite{Porto2019,Porto2021,Porto2022,Porto2023}, where novel PIC algorithms potentially enabling large time steps were exploited to conduct 1D and 2D simulations for an ECR thruster with a magnetic nozzle. In particular, in Ref.\,\onlinecite{Porto2023}, it was explicitly demonstrated that the electron power absorption indeed occurs in the Doppler-broadened resonance zone, where electrons gain perpendicular energy. This leads to electron distribution anisotropy in velocity space, which was argued to be potentially important for certain instabilities. Surprisingly, a second peak far from the ECR region was observed on the profile of the electron mean perpendicular energy. The latter finding was explained by the existence of a large population of electrons trapped by the mirror effect on the back-plate side and by the electrostatic potential on the downstream side of the ECR region.

Another set of works\cite{Takao2014,Takao2016,Hiramoto2017,Nakamura2018,Nakamura2019,Sato2019,Yamashita2019} studied numerically ECR discharges with a grid extraction system. They used the aforementioned simplified approach, where the MW fields were calculated from Eqs. (\ref{EMPIC_eq2}) and (\ref{EMPIC_eq3}), but without taking into account the plasma contribution, i.e., with $\bm{j}=0$. This is not justified for dense or over-dense plasmas, where plasma frequency is comparable to or larger than the driving frequency, and by dropping the plasma influence, one can oversee additional cutoff or resonance locations. For example, Fig.\ref{fig: ECR thruster properties} taken from a PIC simulation, performed in Ref.\,\onlinecite{Yamashita2019} for an ECR discharge driven at $4.25$ GHz, shows that the electron density is close to the critical density, which is approximately $9.8\times 10^{16}$ m$^{-3}$ for this frequency. However, if one is interested in other effects than the EM wave propagation, the exact details of power absorption, and the related electron heating or energization, then such an approximation may be a reasonable assumption. Ref.\,\onlinecite{Takao2014} simulated the $\mu$1 miniature ECR gridded thruster and confirmed that electrons there are well confined due to the mirror effect, which enables their efficient heating. The ions were expected to be well accelerated by the grid system without significant ion losses to the walls. Ref.\,\onlinecite{Takao2016} considered the electron extraction in such a device and concluded that the electric field at the orifice edges played an important role. In Ref.\,\onlinecite{Hiramoto2017}, an azimuthally rotating instability in the discharge chamber was observed, which breaks the azimuthal symmetry and produces an azimuthal electric field, causing an enhanced axial transport of electrons due to the ${\bm E}\times {\bm B}$ electron drift with $B$ the magnetic field with a finite radial component. Unlike the previous works considering ECR discharge in Xe, Ref.\,\onlinecite{Nakamura2018} simulated a similar discharge in water vapor, albeit accounting only for a few positive ion species, and found that in order to obtain the same electron density, a higher power is needed compared to discharges in Xe (see Sec.\, \ref{subsec: alternative propellants} for details). It was also seen that H$_2$O$^+$ and OH$^+$ dominated over H$^+$, and occupied more than 97\%. 
Despite this, the H$^+$ ions accounted for about 10\% of the ion current density due to their small mass, so these ions should not be omitted. An azimuthal instability seen before \cite{Hiramoto2017} was detected. Ref.\,\onlinecite{Nakamura2019} went further and added the H$^-$, O$^-$, and OH$^-$ negative ion species to the previous model. The results indicated that, although the H$_2$O$^+$ still dominated, one saw the emergence of the large-scale spoke structures resembling those observed in planar magnetrons with a similar magnetic field geometry\cite{Hecimovic2018}. The spokes were found to have a negative effect on the truster performance as they produced an electron backflow. Ref.\,\onlinecite{Sato2019} examined the influence of different orifice and magnetic field shapes on the extraction efficiency in a similar discharge but operated with xenon. It was found that, whereas the orifice shapes had little impact, one of the considered magnetic field forms allowed to increase the efficiency by 50\%, which was attributed to the backflow reduction due to the absence of spokes and decreased electron losses toward both the downstream inside surface and the outside wall of the discharge chamber. In Ref.\,\onlinecite{Yamashita2019}, basic plasma properties were obtained from simulations of $\mu$10 thruster operated in xenon. Fig.\,\ref{fig: ECR thruster properties} demonstrates some of the resulting data, where the dashed curve indicates the ECR contour. It can be seen that both the electron temperature and the ionization rate peak close to this contour. This is to be expected for the simplified model, which neglected the plasma contribution, but it might be modified if a more general model is used. The electron density profile is much broader than that of the ionization rate, which indicates some diffusion across the magnetic field, but has an arc-shaped boundary, suggesting that the electrons were well confined by the magnetic field. It was additionally found that another important ingredient to electron confinement was the electrostatic potential and that consequently the confinement depends on the electron energy. This and another finding that the EEDF is anisotropic in velocity space were later corroborated with a different model\cite{Porto2023}. The energetic electron population was shown to be a major player in ionization, whereas its density was relatively small. It was also noted that the magnetic field influence on ions could be important. Another group used a similar, albeit two-dimensional, model for the field-plasma dynamics to model another ECR gridded thruster setup \cite{Fu2019,Fu2021,Fu2022}, where, in contrast to previously mentioned models, Coulomb collisions were included. Ref.\,\onlinecite{Fu2019} pointed out the importance of the effects related to magnetic field non-uniformities. Ref.\,\onlinecite{Fu2021} considered electron extraction mechanisms and identified two different electron extraction channels in the ECR neutralizer. In the first channel, extracted electrons moved across the magnetic field lines starting from the ECR region, where most of electrons were confined by the mirror effect and were heated by the ECR mechanism. In the second channel, extracted electrons moved along the magnetic field lines starting from the periphery of the ECR region. It was observed that the electron transport in these channels had different response to an increasing anode potential. 
In Ref.\,\onlinecite{Fu2022}, the neutral gas dynamics were modeled self-consistently with the field-plasma evolution using the DSMC approach augmented with an adaptive particle management algorithm.

\paragraph{Helicon plasma thrusters \label{HPT}}
Helicon discharges have a number of features, which makes them attractive for EP applications\cite{Takahashi2019,Takahashi2020}. They have high ionization \cite{Boswell1984} and power conversion efficiencies \cite{Takahashi2022}, the latter ensured by the absorption of the helicon and the Trivelpiece-Gould (TG) modes excited in a plasma with a relatively small magnetic field \cite{Chen1991}, which is sustained by either a solenoid or a permanent magnet. The excitation frequency lies in the range $\omega_\mathrm{c,i}\ll \omega \ll \omega_\mathrm{c,e}$, which translates for typical parameters to MHz and thus does not require complicated power generators. Conceptually, it can be viewed as an ICP discharge enhanced with a DC magnetic field. Therefore, they share some of the features inherent to ICPs, such as the electrodeless configuration and the E-H heating mode transition with increasing power. However, they also feature a unique physics when the power is further increased, the helicon and TG modes are excited, and the discharge goes into the W heating mode \cite{Isayama2018}.

There is a rich physics anticipated for a helicon discharge in the W heating mode, which would require full EM PIC simulations. The excitation of helicon and TG modes, their absorption in the plasma, and the related non-linear physics, such as the mode conversion, parametric instabilities, non-linear Landau damping, and non-linear development of various instabilities are of particular interest in this context\cite{Shinohara2022,Chen2006,Isayama2018,Shamrai1997,Virko2003,Niemi2008}. The power absorption and electron heating profiles are intertwined with many other important phenomena taking place in helicon plasmas, such as the shaping of the plasma density profiles \cite{Emoto2023} and the triggering of various instabilities, leading to the anomalous electron transport across the magnetic field lines \cite{Isayama2018}. Due to the low collisionality, the EEDFs are often non-Maxwellian, and the mean free path can be comparable to the system size.
Hence, the corresponding simulations must be self-consistent, kinetic, and non-local so that PIC simulations suit ideally. Unfortunately, due to the high plasma density generated in helicon plasmas in this regime, the corresponding Debye length is small, and due to the limitation on the cell size and the time step for the conventional explicit PIC algorithm, the corresponding computational cost is high. This is why PIC simulations of helicon discharges available in the literature consider the low-power regimes, where plasma density is not very large and, possibly, power absorption dynamics is similar to the ICP regime and can be modeled as such (see ``Inductively coupled thrusters''). In other cases, when the focus is not on wave propagation and self-consistent power absorption, one can model certain aspects of interest with an electrostatic PIC. Another option is to use a hybrid code, where the electron response is accounted for using a fluid approach and generally neglecting the inertia effects, with the ion dynamics modeled with the PIC method \cite{Zhou2019}. Such an approach would be limited by the stiffness on the ion time scale and can be applied to study selected problems, but is at risk of omitting an important physics related to the electron inertia (which can be important in the frequency range around the lower-hybrid frequency that helicon thrusters are commonly operated in), non-local, and electron kinetic effects. 

Due to the aforementioned complexity of the EM PIC modeling, the reports of corresponding fully self-consistent EM PIC simulations of helicon discharges are scarce in the literature\cite{Olshansky2018,Jaafarian2018}. Ref.\,\onlinecite{Olshansky2018} considered the H-W heating mode transition as the magnetic field and the power were varied. The results indicated that, in contrast to a common assumption that in the W regime the power absorption is dominated by the TG modes, it was not so for the considered discharge. Instead, the power was absorbed by electrons predominantly in the plasma bulk, stressing the importance of using self-consistent EM PIC models. Ref.\,\onlinecite{Jaafarian2018} also suggested that the helicon modes can be significant for the description of power absorption. 

Other works used a simplified description of the electromagnetic field and, as a result, power absorption, where it was assumed that the discharge was operated in the low-power H regime so that the ICP description of the $\bm{E}_\mathrm{EM}$ applied (see ``Inductively coupled thrusters''), but focused on other aspects of the discharge physics\cite{Takase2018,Emoto2021,Emoto2021a,Emoto2023}. Ref.\,\onlinecite{Takase2018} modeled neutral dynamics using the DSMC method and considered injection of neutrals from upstream and downstream sides, as long as a magnetic field variation. The output data showed that the downstream injection combined with the magnetic field being strongest close to the thruster exit led to a shift of the plasma density peak from the upstream to the downstream side, which resulted in a larger total thrust. Ref.\,\onlinecite{Emoto2021} examined electron and ion momentum gain in magnetic nozzle acceleration, finding that the axial momentum gain of electrons increased significantly with increasing magnetic field strength becoming dominant in the magnetic nozzle and that the axial momentum gain of electrons was caused by the electron momentum conversion from the radial to the axial direction, resulting in a significant increase of both thrust and specific impulse. In Ref.\,\onlinecite{Emoto2021a}, it was discovered that with an increasing magnetic field, the contribution of the diamagnetic current prevailed over the $\bm{E} \times \bm{B}$ current for what regards the thrust generation. Ref.\,\onlinecite{Emoto2023} investigated the change in plasma density profile and the transport of energetic electrons with increasing magnetic field. It was observed that the center-peaked density profile becomes bimodal, and it was suggested that electrons were heated by the RF electric field and then transported radially inward, leading to a non-local mechanism shaping the ionization rate and plasma density profiles.

\subsection{\label{subsec: plasma wall interaction}	Plasma-wall interaction }
Plasma-wall interaction is a fundamental process in almost all the EP devices characterized by a high surface-to-volume ratio. Many experimental evidences \cite{Gascon03,Barral03,Raitses2006,Tsikata2017} have shown how the thruster wall material has a great impact on thruster performances and discharge parameters. Particle models are well suited for simulating the non-neutral character of the plasma-wall transition region.
The different topics related to plasma-wall interaction and captured by particle-based models can be classified in the following categories: i) electron deviation from a Maxwellian distribution due to sheath effects, ii) electron-induced secondary electron emission, iii) ion sputtering, iv) ion recombination at the walls and v) gas-wall interaction. All these aspects are presented below.

\paragraph{Sheath models \label{sheath}}
Recently, a renewed interest in the study of electron kinetics has been highlighted with a series of papers\cite{Dominguez2018, Dominguez2019, Marin2021, Marin2022} studying the dynamics perpendicular to the lateral walls in HT discharges by means of a revised version of the PIC model of Taccogna \textit{et al.}\cite{Taccogna2008,Taccogna2014}. Since the low collisionality is insufficient to replenish the high-velocity electrons collected at the walls, a significant depletion of the parallel-to-magnetic field electron velocity distribution function (VDF) is detected that has several important implications on some global quantities: sheath potential drop, wall collision frequency, particle and power wall losses \cite{Kaganovich07}. In particular, the VDF depletion permits to have  quite lower electron particle and energy fluxes reaching the wall, compared to the Maxwellian population, reducing the impact of the near-wall conductivity on the total anomalous electron cross-field transport. Moreover, the low electron collisionality introduces an anisotropy between perpendicular and parallel to B field electron temperatures, and asymmetries between the inner and outer walls in HT discharges. However, it has been shown that the magnetic field inclination relative to the lateral walls affects the transfer between radial and axial electron velocity components and reduces these effects.

\begin{figure}
    \centering
    \includegraphics[scale=1.0]{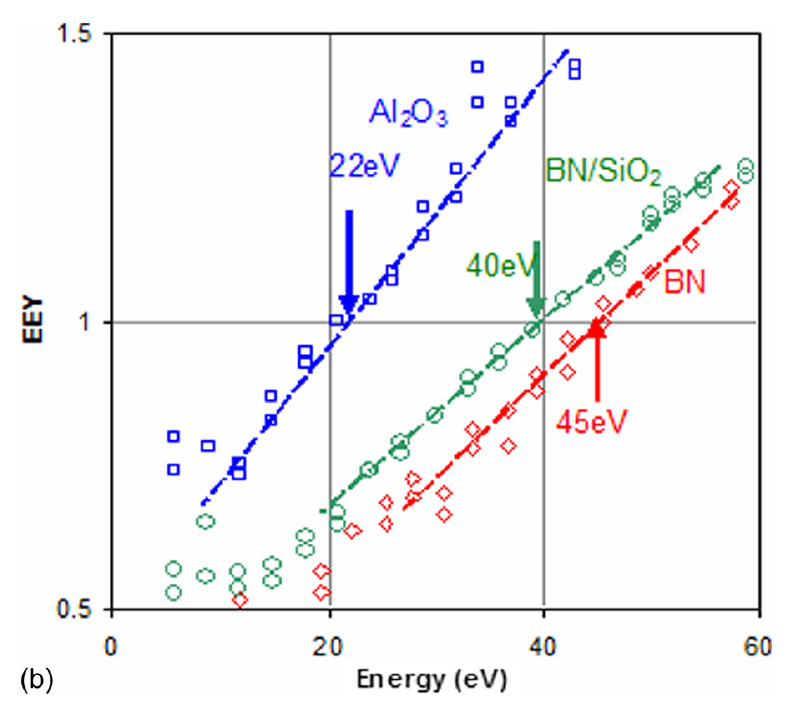}
    \caption{Electron-emission yield for pristine $Al_2O_3$, $BN$, and $BNSiO_2$. The lines corresponds to linear laws. The arrows show the first crossover energy. Reproduced with permission from Tondu \textit{et al.}\cite{Tondu11}, Journal of Applied Physics \textbf{110}, 093301 (2011). Copyright 2011 American Institute of Physics.}
    \label{fig: SEE}
\end{figure}

\begin{figure*}
    \centering
    \includegraphics[scale=0.115]{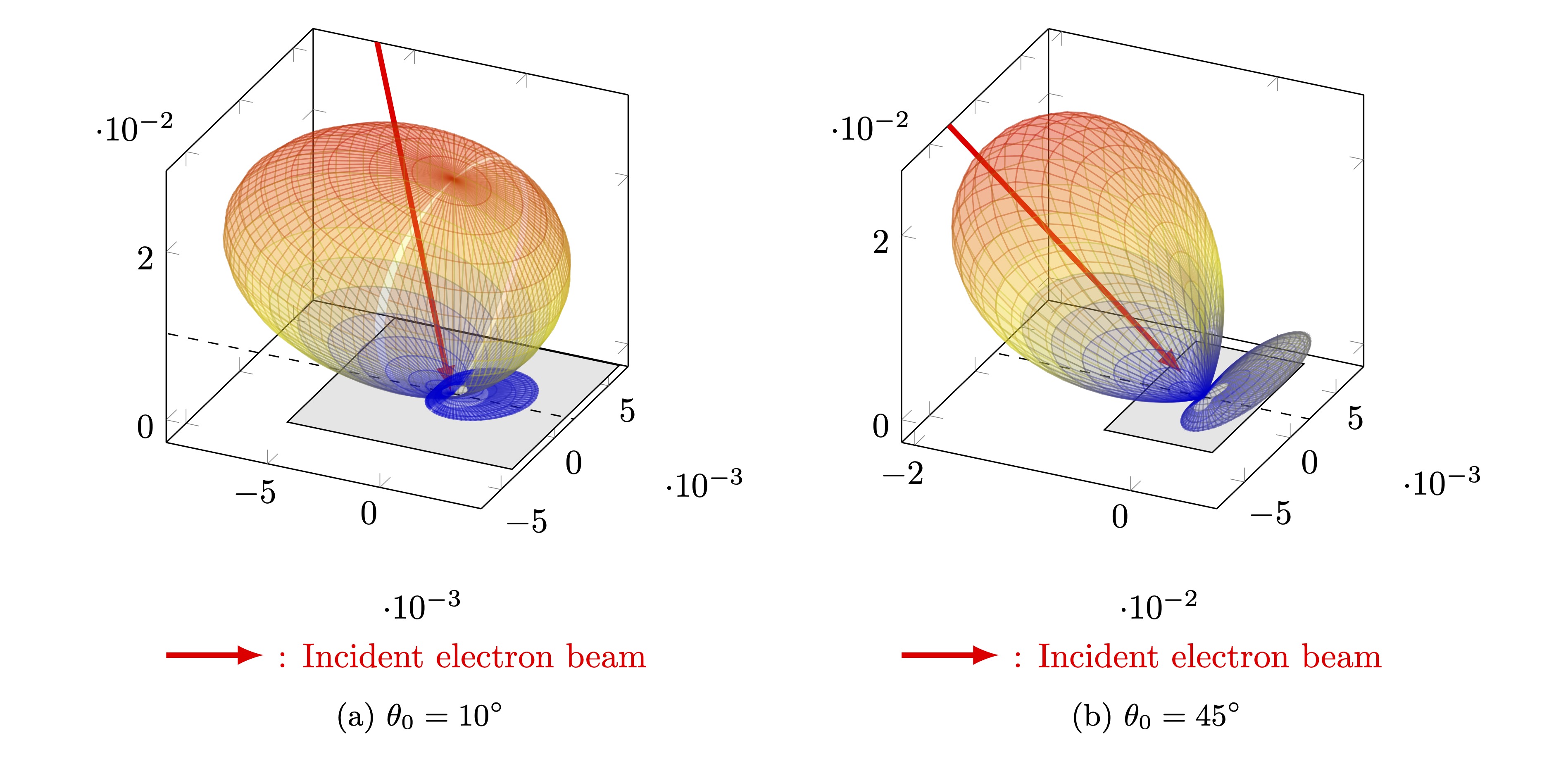}
    \caption{Elastic backscattering lobes for Al surface with an incident electron energy $E_\mathrm{p}$ = 40 eV and two different incident angle: (a) $\theta_\mathrm{p}$ = 10° and (b) $\theta_\mathrm{p}$ = 45°. Reproduced with permission from Villemant \textit{et al.}\cite{Villemant17}, IEPC-2017-366 (2017), Copyright 2017 Electric Rocket Propulsion Society.}
    \label{fig: theta_SEE}
\end{figure*}

\paragraph{Secondary electron emission models \label{sEE}}
The secondary electrons emission (SEE) is mainly caused by electron impacts (with a negligible influence of ion impacts) when the lateral surfaces of the thruster are made of a dielectric material, which is a quite common case. 
Electron-induced SEE has an important effect on the lateral potential sheath drop, the wall energy losses, the absorption power and the electron cross field transport (near wall conductivity) and it is generally coupled with instabilities \cite{Taccogna09}.
Due to the typical fast timescale characterizing the electron-material process ($10^{-13}$ sec), the electron wall emission implementation in PIC models can be done by a phenomenological approach. In the last years, different SEE algorithms have been proposed that are characterized by the electron emission yield (EEY) $\sigma(E_\mathrm{p},\theta_\mathrm{p})$ and the spectrum of emission energy $f_\mathrm{E}(E_\mathrm{p},\theta_\mathrm{p})$ and angle $f_\theta(E_\mathrm{p},\theta_\mathrm{p})$ of secondary electrons, all functions of the impact energy $E_\mathrm{p}$ and angle $\theta_\mathrm{p}$ of the primary electron: the linear and power law \cite{Barral03,Dunaevsky03}, the modified power law \cite{Tolias14}, the modified Vaughan model \cite{SydorenkoPhD06}, the Sombrin model \cite{Villemant19}, the Furman-Pivi model \cite{TaccognaPhD03} and models using expressions obtained from machine learning software \cite{Chang19}.
However, there is still lack of data related to the EEY, and especially in the low-energy range $E_\mathrm{p} <$ 30 eV (corresponding to the impact energy of a large fraction of primary electrons due to the decelerating effect of the sheath electric field), owing to experimental difficulties to measure it. The most accurate measurements \cite{Tondu11,Andronov13,Gonzalez17,Loffler2023} (see Fig.\,\ref{fig: SEE}) and theoretical works \cite{Bronold20} suggest a non-zero EEY at zero impact energy $\sigma(E_\mathrm{p}=0) > $ 0.3-0.4 with even an increasing EEY for decreasing energy $E_\mathrm{p}$ lower than 10 eV for most common dielectric materials relevant to EP. This behavior is not reproducible with linear, power or Vaughan laws and can be ascribed to the different behaviour of the EEY of the three populations of emitted secondary electrons: elastically and inelastically backscattered, and true secondaries (electrons belonging to the wall material).
Some numerical works \cite{Pigeon20} have shown how the macroscopic behavior can change according to the value assumed by the EEY at $E_\mathrm{p}=0$.
Finally, backscattering electrons have a memory effect of the impact energy and angle \cite{Jablonski14,Villemant17}: while true secondaries have an isotropic emission (cosine-Lambertian distribution \cite{Greenwood02,Costin20}), backscattered electrons show a double-lobe angular emission \cite{Petillo02,Villemant17,Gueroult18} corresponding to the almost incident and specular angles (see Fig. \ref{fig: theta_SEE}).
This can have important consequences on the non-local character (electrons emitted from one wall are often those impacting on the opposite wall) and on the realistic estimation of the near-wall contribution on the electron anomalous mobility.

\paragraph{Ion-wall interaction models \label{IWI}}
The lifetime of different electric thrusters is limited by the large erosion of the chamber walls due to ion sputtering: in magnetic unshielded configurations, the integrated ion flux to the walls can represent up to 40\% of the total ion production (integrated ionization source term over the thruster volume).
Therefore, a reliable and precise simulation of the discharge wall erosion would be beneficial to reduce long and expensive life tests in vacuum chambers. These simulations require to know both the ion velocity distribution function at wall impact (this can be obtained using a particle model at least for ions) and the sputtering yield function of the wall material in terms of impact energy and angle (for a given impacting ion species). 
Computational efforts have successfully reproduced general erosion rate trends \cite{cho13}, but models are not yet fully predictive or capable of reproducing experimentally observed surface features. In particular, the low ion energy sputtering yield is still not well know, the connection between erosion and performance degradation requires further study, and an explanation of the ubiquitous presence of the so-called ``anomalous” erosion ridges in HT configurations remains elusive \cite{Brown2020}. High-fidelity plasma models have yet to be integrated with sophisticated material and sputtering models. Finally, a complete assessment of the erosion effects requires to simulate the re-deposition rates, and therefore to follow the sputtered atoms trajectories. This is generally less expensive computationally than assessing the ion sputtering profiles, as these neutrals trajectories are weakly coupled with the plasma and can be studied using simplified approaches, like the view factor models \cite{arak23}.

When ions hit the thruster walls, apart from possibly causing the emission of sputtered atoms, they most likely lose their kinetic energy and tend to recombine with wall electrons. This process is known as `` ion recombination'', and, for a saturated wall, nearly all impacting ions eventually return to the plasma as neutrals. Actually, impacting ions can also be reflected by the wall \cite{ecks86}, especially at grazing incidence angles and for very low mass ratios between the ion and the wall atoms, being the reflection probability negligible above a mass ratio of 2. Since ions for plasma propulsion are relatively heavy and significantly accelerated towards the wall inside the plasma sheath (hence they feature a close-to-normal incidence angle), ions reflection is typically neglected in most particle codes for plasma thruster simulations.
Ions can finally be implanted into the surface, although this is extremely unlikely for heavy ions to occur, with recent studies showing that for Xe ions against an Al target, the implantation probability is around 0.2-0.5\% at normal incidence and energies above 300 eV \cite{ito21}.
When the ion is neither reflected nor implanted, recombination takes place (vast majority of cases), and the resulting neutral atom can either be adsorbed by the surface (to recombine into a neutral molecule, if this is possible), or be emitted back into the plasma. The probability for molecular recombination in the case of relevant alternative ion propellants such as N and O (see Sec.\,\ref{subsec: alternative propellants}), is respectively 7 and 17\% \cite{Taccogna2022}. While recombined molecules are re-emitted in thermal equilibrium with the wall (i.e. with a mean emission energy equal to $2 T_\mathrm{w}$, where $T_\mathrm{w}$ is the wall temperature in energy units), and with an angular profile correctly represented by a Lambertian-cosine distribution, recombined atoms can be emitted with an average kinetic energy $\left<E_\mathrm{emi}\right>$ that also depends on the impacting ion energy $E_\mathrm{imp}$, as dictated by the energy accommodation coefficient $\alpha_\mathrm{w}$:
\begin{equation}
\label{eq: energy accommodation}
\left<E_\mathrm{emi}\right> = \left( 1 - \alpha_\mathrm{w} \right) E_\mathrm{imp} + 2 \alpha_\mathrm{w} T_\mathrm{w}.
\end{equation}
This coefficient generally depends on several factors, such as the mass ratio between ions and wall atoms, the surface cleanliness and roughness, and both the impacting particle energy and direction. However, in the absence of experimental data (which is generally available for light species at high energy levels \cite{Shuvalov1983,greg86,moe05}), numerical simulations generally assume values close to unity, as suggested also by a recent experimental evidence \cite{liu20}. A large influence on plasma discharge properties, such as the propellant utilization efficiency $\eta_\mathrm{m}$, has been observed in recent parametric studies \cite{gibs17,Guerrero21,domi21}. In particular, a lower value of $\alpha_\mathrm{w}$ produces a more energetic neutrals population, which is less easily ionized inside the thruster, thus reducing $\eta_\mathrm{m}$. Regarding the angular distribution of the emitted recombined atoms, this can deviate from a Lambertian-cosine distribution, especially at grazing ion incidence angles, although the near totality of simulation codes neglect this for the sake of simplicity and for the lack of reliable angular data at the impact energies of interest. Finally, another source of confusion to take into account is that the energy accommodation coefficients and angle distributions reported in literature generally capture the behavior of all reflected particles, including both reflected ions and recombined neutrals and are therefore incompatible with a simultaneous use of an ion reflection coefficient (to model direct ion reflection without recombination).

\begin{figure}
    \centering
    \includegraphics[scale=0.6]{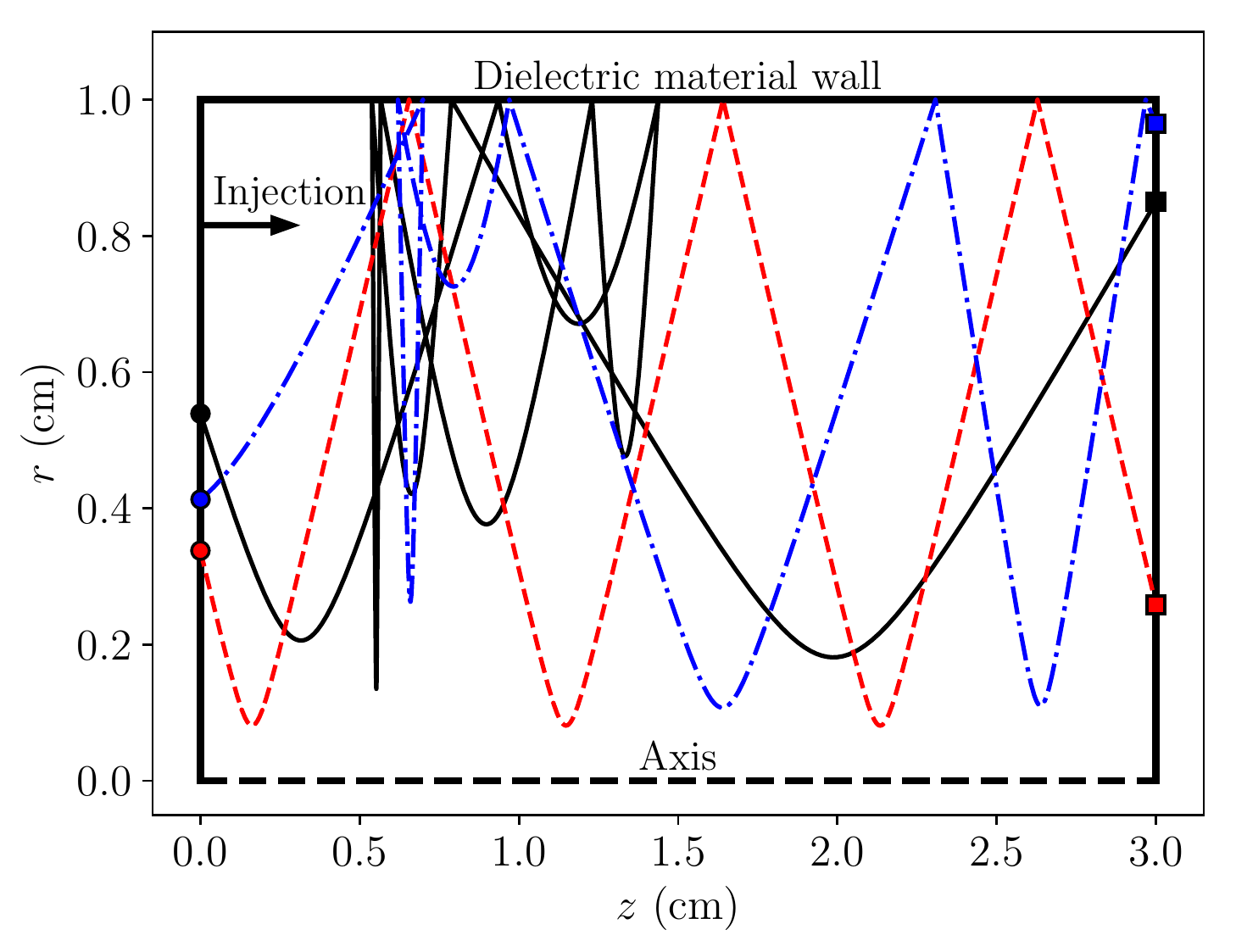}
    \caption{Trajectories of neutrals inside a cylindrical discharge chamber. Propellant atoms are injected from the left boundary, and get ionized by an isothermal electrons fluid. The dashed red line refers to the trajectory of a specularly reflected neutral, the black solid line to a purely diffused neutral, and the blue dash-dot line to a neutral reflected according to Schamberg's model. Reproduced with permission from Domínguez-Vázquez \textit{et al.}\cite{domi21}, Plasma Sources Science and Technology \textbf{30}, 085004 (2021), Plasma Sources Science and Technology. Copyright 2021 IOP Publishing Ltd.}
    \label{fig: neutral reflection effects}
\end{figure}

\paragraph{Gas-wall interaction models \label{GWI}}
Closely related to the ions-wall interaction, the gas-wall interaction is finally another topic of great interest for the simulation of electric thrusters and intake for atmospheric breathing electric propulsion. Although neutrals are not affected by electric and magnetic fields, their density profiles are strongly coupled with those of the plasma, as they determine the ionization source term, and both the momentum and energy loss terms for ions and electrons. In fact, the neutral propellant density can be up to 10 times larger than the plasma density, so that its accurate prediction is very relevant. Propellant neutrals generally feature a kinetic energy of fractions of eV, and, at these energy levels, there is a lack of available data regarding both the energy accommodation of wall-reflected neutrals \cite{Gueroult18} (also modeled with Eq.\,\eqref{eq: energy accommodation}) and their angular distributions. In particular, both experimental data and theoretical models do not cover the low-energy interval of interest. Regarding the former, it is extremely challenging to obtain a mono-energetic beam of slow neutrals, while regarding the latter, at low energies, the particle-wave nature of the impacting particle starts to emerge and classical mechanics models, like the hard cube or soft cube models \cite{loga68}, start to fail. In any case, just like for ions, both the surface roughness and the impact angle affect both the energy accommodation coefficient and the angular distribution of the reflected neutrals. The most commonly used models for the simulation of neutrals reflection are the Maxwell's model, which considers a fixed probability for specular and purely diffused reflections, and Schamberg's model \cite{prie14, domi21}, which depends on additional tuning coefficients to reproduce an intermediate reflection scenario affected by the impact angle. Ref.\,\onlinecite{domi21} reports a study on the effect of the neutrals reflection model in an plasma discharge within a cylindrical chamber, assuming elastic (i.e. energy conserving) collisions with the wall. Fig.\,\ref{fig: neutral reflection effects} shows how this model affects the trajectories and hence the residence time of neutrals inside the chamber, which is clearly overestimated by assuming a purely diffuse reflection. The considered model is observed to significantly affect both the minimum mass flow required for a sustained plasma discharge, and the propellant utilization efficiency $\eta_\mathrm{m}$, at low values of this efficiency (less than 70\%). In fact, at higher $\eta_\mathrm{m}$, neutrals get nearly completely ionized inside the chamber regardless of the reflection model.

The gas-wall interaction is also particularly important for the optimization of the propellant injection location and of the air breathing electric propulsion (ABEP) intake performances. The first study has been recently conducted with a Direct Simulation Monte Carlo (DSMC) model of the Xe gas propellant in a HT \cite{Boccelli21}. The work indicates that a reversed injection (propellant injected from the exit plane backwards towards the anode) yields an increase of the propellant utilization efficiency $\eta_\mathrm{m}$ by a maximum of $30\%$ with respect to the direct injection from the anode. The latter has been investigated by different DSMC models \cite{Romano21,Rapisarda23} that show the effect of flow misalignment and thermal accommodation coefficient $\alpha_W$ on the intake performance. The importance of chemical reactions and recombination of atomic oxygen into $O_2$ molecules on the wall has also been highlighted in order to model the variation of the gas composition throughout the intake \cite{Rapisarda23}.

\section{\label{sec: plumes and SC interaction} PIC models of plasma plume expansion and interaction with the spacecraft}
All electric thrusters produce plasma plumes that expand into free space and can interact with sensitive spacecraft surfaces and with the onboard telecommunications system. Regarding the former, since not all propellant is ionized inside the thruster, the fast emitted ions interact with slow neutral particles through ``charge exchange'' collisions, which have the effect of producing slow ions whose trajectories are subject to the local electric fields. These collisions can therefore produce the so-called ``ion backflow'' towards satellite surfaces such as the solar arrays or optical sensors, whose performance can be significantly degraded by the induced sputtering and contamination/deposition. For what concerns the interference of the electric propulsion subsystem with the telecommunications subsystem, this can be produced by either the thruster antennas/coils (e.g. the RF coil or the Helicon antenna in the corresponding thrusters) or by the plasma plumes themselves, which are not transparent to radio-waves with a frequency below the plasma frequency, a fact that generally occurs in the most dense plasma regions, close to the thruster.

For the above considerations, the system engineer of the hosting satellite must carefully select the installation position of the electric thruster, with the help of accurate plasma plumes simulations, which should predict the plasma properties in the surroundings of both the thruster and the spacecraft, thus covering distances up to several meters. Currently, there exist two main simulation approaches to tackle this challenging task: the hybrid approach and the full PIC approach, which are presented and compared in the following sub-sections.

\subsection{Hybrid models and their limitations}
In order to simulate the plasma plume interaction with a generic spacecraft of arbitrary geometry, the simulation model must necessarily be 3D, with the consequent computational burden. Hybrid models \cite{Taccogna02,Taccogna04,Taccogna14b,ahed23,cich17,cai15,kork15,fill21,kang23,arak19} have therefore emerged in this context, as they represent the best compromise in terms of accuracy and computational cost for modeling the plasma thruster plume expansion and its interaction with the spacecraft. In fact, they feature a kinetic treatment of the ion species (and in some cases of the neutrals as well) so that CEX collisions are correctly captured, and a fluid modeling of the fastest species, the electrons, which are subject to conservation equations. These features permit to avoid both time and spatial constraints of a full PIC simulation: the electron CFL condition or the plasma and cyclotron frequency constraints for what concerns the time step, and the Debye length constraint in quasi-neutral plume regions for what regards the cell size. 
The complexity of the electron fluid model depends on the considered closure for the electron conservation equations, which can be either a pressure tensor closure \cite{cich17, fill21, kang23, arak19}, or a heat flux vector closure \cite{ahed23, cai15}. 

In the first case, a commonly made choice is to assume polytropic electrons, whose temperature is a function of the electron density: $T_\mathrm{e} (n_\mathrm{e}) \propto n_\mathrm{e}^{\gamma-1}$. Hybrid models belonging to this category can be further distinguished on the basis of the considered terms in the momentum balance equation. The simplest approach is to neglect all inertial, collisional, and magnetic field effects \cite{Taccogna02,Taccogna14b,meri15, fill21} and obtain the electric potential from Boltzmann's relation as
\begin{align}
    \begin{cases}
    \phi = \dfrac{T_\mathrm{e0}} {e} \ln \left( \dfrac{ n_\mathrm{e}}{n_\mathrm{e0}} \right) & \,\, \mathrm{for}  \,\, \gamma = 1 \\
    \phi = \dfrac{\gamma T_\mathrm{e0}}{e(\gamma-1)} \left[ \left( \dfrac{n_\mathrm{e}}{n_\mathrm{e0}}\right)^{\gamma-1} - 1\right] & \,\, \mathrm{for} \,\, \gamma \ge 1,
    \end{cases}
\end{align}
where $T_\mathrm{e0}, n_\mathrm{e0}$ are the reference electron temperature and density at a point where $\phi=0$. The asymptotic electric potential (i.e. the potential limit for $n_\mathrm{e} \to 0$) is therefore $-\infty$ for isothermal models ($\gamma=1$), a clearly unphysical prediction, and a more realistic $- \gamma T_\mathrm{e0}/\left[ e (\gamma-1) \right]$ for $\gamma>1$. Other more complex models obtain $\phi$ (or a related thermalized potential $\Phi$) by solving the electron momentum balance equation, accounting for collisional effects \cite{cich17}, which permit to obtain the electron current density, electron inertia effects \cite{pera21}, relevant in certain regions characterized by large plasma gradients in the very near field plume, and magnetic field effects \cite{cich20, cich22}, which are omnipresent in plasma plumes expanding in low Earth orbit (due to the geomagnetic field) or very relevant in the near-field plume region of HT and most EM thrusters. The main drawback of these models based on polytropic electrons, however, is that they fail to reproduce correctly the electron temperature in magnetized plumes from HT \cite{cich21}, but also in unmagnetized scenarios, as shown later.

In the second case, the heat flux closure, the electron temperature is obtained by solving an electron energy conservation equation \cite{ahed23, cai15}. Different types of closures have been considered: diffusive Fourier-like laws, with $\bm{q}_\mathrm{e} \propto -\nabla T_\mathrm{e}$, but also hybrid diffusive-convective closures, in which the electron heat flux vector has a component that is proportional to the electron fluid velocity. In this respect, it has been shown \cite{ahed20} that a purely convective law for the heat flux vector, $\bm{q}_\mathrm{e} = \alpha n_\mathrm{e} T_\mathrm{e} \bm{u}_\mathrm{e}$ is equivalent to a polytropic electron model, provided that $\gamma = \left(5 + 2\alpha \right)/\left(3 + 2\alpha \right)$.

\begin{figure}
    \centering
    \includegraphics[width=0.5\textwidth]{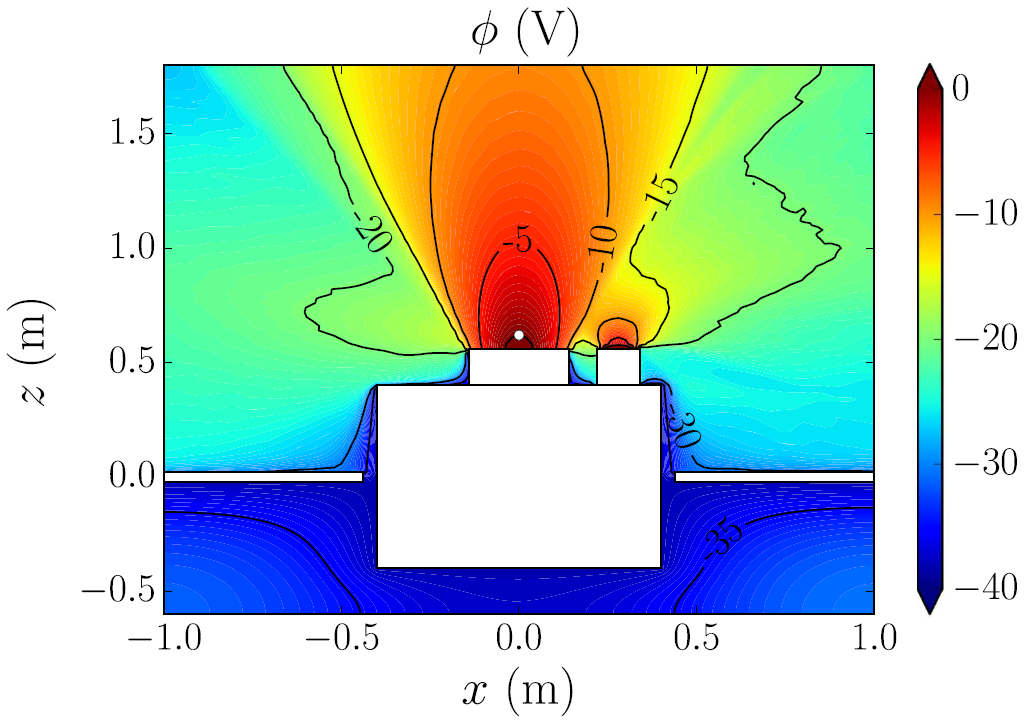}
    \caption{Electric potential in the surrounding of a cubic satellite, predicted by a hybrid model, with polytropic electrons modeled with $\gamma=1.1$ and non-neutral regions taken into account. Charge-exchange collisions for ions are included. Adapted with permission from Cichocki \textit{et al.},\cite{cich17}, Plasma Sources Science and Technology \textbf{26}, 125008 (2017). Copyright 2017 IOP Publishing.}
    \label{fig: hybrid model results}
\end{figure}

Hybrid models can also include non-neutrality effects, which are very relevant close to the spacecraft surfaces or in the peripheral plume regions, by accounting for Poisson's equation. In models featuring a pressure tensor closure, this can be added as a non-linear Poisson's equation to correct both the electric potential and the electron density in regions where the quasi-neutrality condition, $\epsilon_0 \nabla^2 \phi^* \ll e n_\mathrm{e}^* $, is violated. Here $\phi^*,n_\mathrm{e}^*$ are the electric potential and electron density obtained by assuming quasi-neutrality. In such regions, it is possible to express the electron density as a function of the electric potential, $n_\mathrm{e}= n_\mathrm{e}(\phi)$ \cite{cich17, cich18, arak19}, thus yielding a non-linear partial differential equation. The simulation results from a hybrid model assuming polytropic electrons and charge non-neutrality are shown in Fig.\,\ref{fig: hybrid model results}, where the inclusion of non-neutral regions close to the spacecraft is clearly visible (the potential adapts to the satellite ground potential) and has the effect of increasing the ion back flow current by approx. 20\%. This non-linear Poisson's solver approach, however, is not considered by models featuring a heat flux closure, where Poisson's equation is rather coupled to the other conservation equations as a correction of the electron density \cite{mode22}: $n_\mathrm{e} = n_\mathrm{e}^* + \left(\epsilon_0/e\right) \nabla \phi^2$. Here, $\phi$ is the electric potential retrieved from the momentum balance equation, where charge non-neutrality has been assumed. 

\begin{figure}[!ht]
    \centering
    \includegraphics[width=0.47\textwidth]{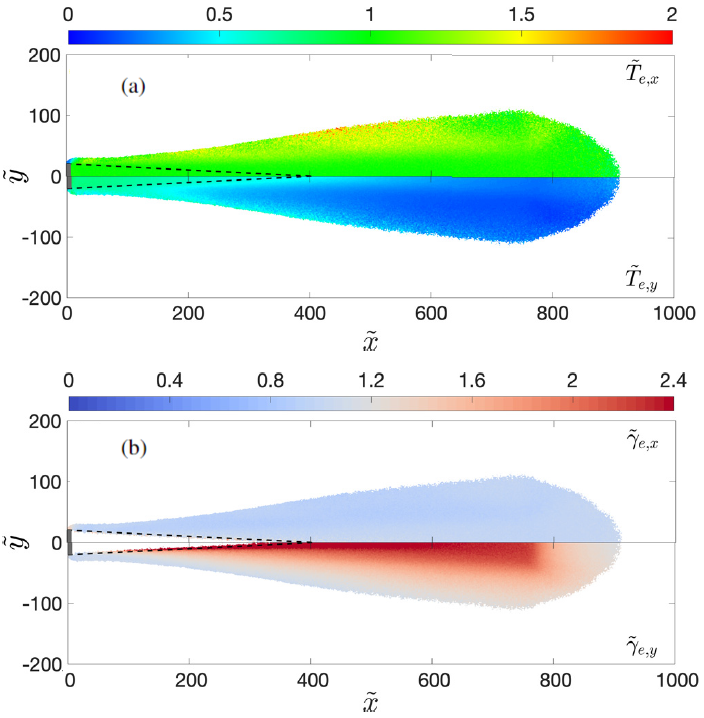}
    \caption{Full-PIC simulation results showing (a) the electron temperature parallel to the plume axis $T_\mathrm{e,\tilde{x}}$ (top) and perpendicular to it, $T_\mathrm{e,\tilde{y}}$ (bottom), and (b) the local polytropic cooling rate for the parallel (top) and perpendicular (bottom) electron temperature. A current-free and quasi-neutral injection from $\tilde{x}=0$ is considered (simulation B of Ref.\,\onlinecite{wang19}). The considered geometry is planar, with the plume extending infinitely in the direction normal to the page. Adapted with permission from Wang \textit{et al.} \cite{wang19}, Physics of Plasmas \textbf{26}, 103502 (2019). Copyright 2019, AIP Publishing.}
    \label{fig: cooling rate and temperature anisotropy with full PIC}
\end{figure}

As mentioned above, however, the polytropic electron thermodynamics (by far, the most common assumption in hybrid plume models) misses very important kinetic effects, as shown by recent studies on unmagnetized plasma plume expansions \cite{meri18, hu17, hu18, wang19, nuwa20}. Merino \textit{et al.} \cite{meri18} have shown, with a kinetic model based on Vlasov's equation, that electrons belong to three main groups: ``free electrons'' that have enough energy to go from the plasma source to infinity, ``reflected electrons'' that eventually get back to the source, and ``doubly trapped electrons''. The contributions of these populations change across the plume expansion, with the reflected and doubly trapped electrons vanishing downstream. As a result, the electron velocity distribution is anisotropic and different cooling rates (varying with the expansion) exist for the electron temperature parallel to the plume axis, and perpendicular to it. While the parallel temperature tends to an asymptotic value, hence with a cooling rate $\gamma_\parallel \to 1$, the perpendicular electron temperature tends to 0, with a cooling rate $\gamma_\perp \to \gamma_{\perp,\infty} > 1$. Recent full PIC studies \cite{hu17, hu18, wang19}, also observed a clear anisotropy in the electron parallel and perpendicular temperatures of a planar plume expansion. Fig.\,\ref{fig: cooling rate and temperature anisotropy with full PIC} shows the evolution of the electron parallel and perpendicular temperatures (subplot (a)), and the corresponding cooling rates (subplot (b)). The parallel electron temperature seems to be nearly constant (at least for the already charged-neutralized plume considered), with a corresponding cooling rate close to 1 and dimly dependent on the distance from the symmetry plane (in this case the plane $y=0$). Instead, the perpendicular electron temperature shows a very quick decay, with a cooling rate that now depends on the symmetry plane distance and that approaches (and even surpasses) the adiabatic limit for the considered planar expansion (with a limit at 2).

\begin{figure}[!ht]
    \centering
    \includegraphics[width=0.47\textwidth]{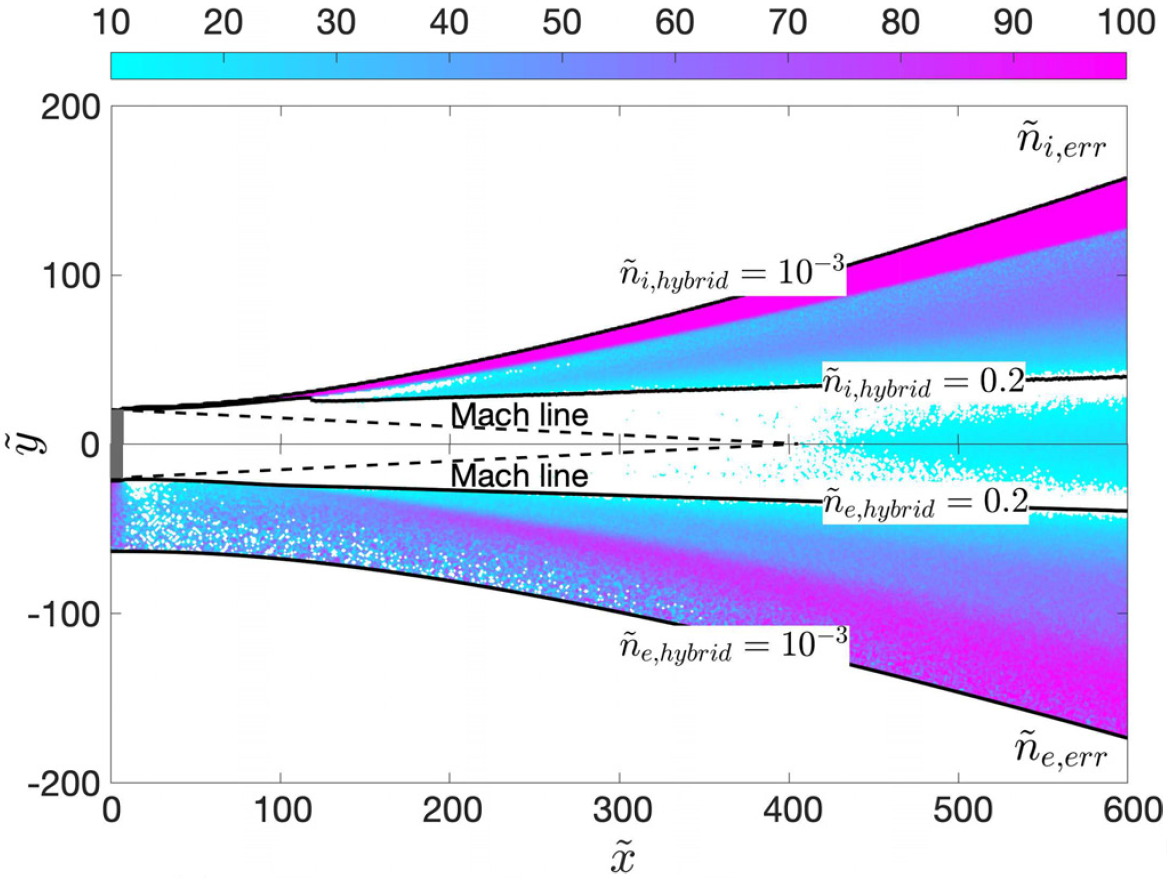}
    \caption{Ion (top) and electron (bottom) relative density differences, between a quasi-neutral hybrid PIC simulation with isothermal Boltzmann's electrons, and a full PIC simulation (simulation B of Ref.\onlinecite{wang19}). Adapted with permission from Wang \textit{et al.} \cite{wang19}, Physics of Plasmas \textbf{26}, 103502 (2019). Copyright 2019, AIP Publishing.}
    \label{fig: density errors Boltzmann versus full PIC}
\end{figure}

It is therefore not surprising that a polytropic electron model based on a single coefficient $\gamma$ cannot capture correctly the above physics and introduces errors even in macroscopic quantities such as the electric potential and the ion density, especially in the plume peripheral and backflow regions. Recent studies \cite{wang19,hu18,nuwa20} have explicitly shown the shortcomings of a simple quasi-neutral Boltzmann electrons model. While part of these differences can also be ascribed to the quasi-neutral assumption considered for the hybrid model, the kinetic effects described above are clearly missed by the latter and have a non-negligible effect in macroscopic plasma plume properties. Fig.\,\ref{fig: density errors Boltzmann versus full PIC} shows the relative differences in ion and electron densities between a Boltzmann model and the same full PIC simulation of Fig.\,\ref{fig: cooling rate and temperature anisotropy with full PIC}: while the density is correctly captured within the plume core, relevant errors are found in the peripheral regions, for both ions and electrons. Finally, Nuwal \textit{et al.} \cite{nuwa20} compared full-PIC and Boltzmann/polytropic electron models, in a spacecraft interaction scenario for a collisional plume with CEX collisions. This study highlighted important differences in the predicted slow ions population (due to CEX), which is the main responsible for spacecraft sputtering, charging and contamination. 

\subsection{PIC models}
\label{subsec: PIC models for plumes}
In full-PIC plasma plume simulations, two critical topics are worth further discussion: the particles loading, and the boundary conditions for both particles and fields. 

Ions and electrons can be loaded into the simulation domain from (i) a quasi-neutral and ambipolar injection surface \cite{wang19, li19}, (ii) from a non-neutral ambipolar injection surface \cite{wang19,nuwa20}, in which a neutralization sheath that accelerates the electrons downstream is simulated, or (iii) from different non co-located surfaces for ions and electrons \cite{jamb20}, for studies on neutralizer-ion beam coupling. Achieving current ambipolarity is straightforward by loading the same number of electrons and ions from the emission surface, and refluxing particles that return to the latter \cite{proc90} (e.g. re-injecting them with the same loading distribution). For a locally current free surface and typical values of the ion Mach number (10-30), the number of electrons crossing the emission surface towards the simulated domain per unit time (due to both loaded and refluxed electrons) may be quite larger than that of the ions due to the very different thermal fluxes. Guaranteeing a quasi-neutral emission surface, on the other hand, is more subtle. One possibility is to either load ions and electrons with the same drift velocity into a volumetric source region \cite{wang19,hu18}, which features a fixed plasma potential and adapts in time to guarantee the emission of a quasi-neutral and current-free plume, or by controlling the amount of emitted electrons at each time step, depending on the observed local charge \cite{li19}.
For what regards the particles loading distributions, ions are generally injected according to a drifting Maxwellian distribution, with a small temperature and a hypersonic fluid velocity ($u_\mathrm{\parallel,i} \gg c_\mathrm{s}$, where $c_\mathrm{s}$ is the ion acoustic speed), representative of the considered thruster and, in some cases, accounting for a local plume divergence angle \cite{li19}. The electrons, on the other hand, are loaded with a Maxwellian distribution in the plane perpendicular to the plume axis, and with either a half Maxwellian \cite{li19, jamb20, nuwa20} or a half Maxwellian-flux distribution \cite{wang19} in the direction of the plume axis, that is:
\begin{align}
\begin{cases}
f_\mathrm{inj} (v_\parallel, v_\perp) \propto \left| v_\parallel \right| \exp \left( -\dfrac{m v^2}{2 T} \right), \,\, \mathrm{for} \,\, v_\parallel > 0 \\
f_\mathrm{inj} (v_\parallel, v_\perp) = 0  \,\, \mathrm{for} \,\, v_\parallel \le 0,
\end{cases}
\label{eq: maxwell flux distribution}
\end{align}
where $v^2 = v_\parallel^2 + v_\perp^2$. As shown in Ref.\,\onlinecite{proc90}, it is this latter distribution that correctly reproduces a half-Maxwellian distribution function at the injection surface. For a globally current free plume, the full-Maxwellian electron distribution at the source is recovered due to the emitted electrons that are reflected back towards the source, with the exception of an empty region at high negative velocities (i.e. returning to the source) due to free electrons lost downstream \cite{li19}.

\begin{figure}[!ht]
    \centering
    \includegraphics[width=0.45\textwidth]{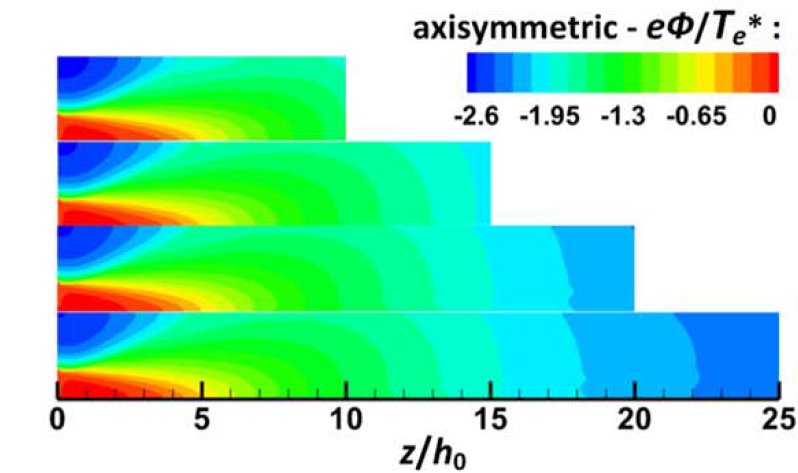}
    \caption{Electric potential in a 2D axisymmetric plasma plume simulation for different domain extensions, and using open boundary conditions for electrons. Reproduced with permission from Li \textit{et al.}\cite{li19}, Plasma Sources Science and Technology \textbf{28}, 034004 (2019). Copyright 2019, IOP Publishing.}
    \label{fig: Open BC effects}
\end{figure}

Coming now to the boundary conditions, for what concerns the electric potential, a Dirichlet condition is generally applied at the injection boundary, and homogeneous Neumann conditions are imposed at the remaining open simulation boundaries. Things are more tricky for what concerns the macro-particles: while ions crossing the simulation boundaries are simply removed from the simulation, electrons are either removed \cite{wang19, hu18}, or selectively reflected \cite{jamb20,li19}. The first approach suffers from more restrictive computational constraints since boundary effects are more relevant and larger simulation boxes are required. This is clear in Fig.\,\ref{fig: cooling rate and temperature anisotropy with full PIC} where results have been analysed before the quasi-neutral plume reaches the downstream boundary. The second approach, on the other hand, is capable of reducing downstream boundary effects by reflecting a fraction of outer boundary-crossing electrons, based on their mechanical energy \cite{jamb20}, or on the global net current exiting the domain in the last PIC steps \cite{li19}. The goal of these approaches is to mimic the partial reflection of electrons taking place further downstream, outside of the simulated domain, and therefore to reduce the influence of the boundaries on the achieved results, as shown in Fig.\,\ref{fig: Open BC effects}.

\subsection{\label{subsec:circuit modeling} Circuital characterization of plasma-spacecraft interaction}

In order to fully characterize the plasma-spacecraft (S/C) interaction, the plasma plume model must be appropriately coupled with a representative circuital model of the satellite surfaces. In fact, the latter affect the plasma, as they impose given electric potentials and/or plasma currents and represent ion recombination sources (into neutrals) and electrons sinks, while the plasma affect the S/C surfaces through both direct sputtering damage and electric currents that can yield to a charge build-up and eventually electrical components breakdown. Past studies have tried to couple a plasma plume model to a circuit model of the spacecraft, or at least to assess the plume induced erosion on the most relevant surfaces \cite{yan12}. In all cases, given the extremely large computational domain and the intrinsic multi-scale character of the plasma-satellite interaction, the plasma plume model considered has been of the hybrid type (typically with polytropic fluid electrons). The circuital representation of the S/C components in contact with the plasma can either consider lumped elements \cite{cich17, cich18} (i.e. the different macro-surfaces are treated as single nodes) or distributed elements with macro-surfaces discretized with a large number of elements or circuit nodes to handle very complex satellite geometries, with the use of unstructured meshes with tetrahedral volume elements \cite{fill21, arak19}. 
\begin{figure}[!ht]
    \centering
    \includegraphics[width=0.50\textwidth]{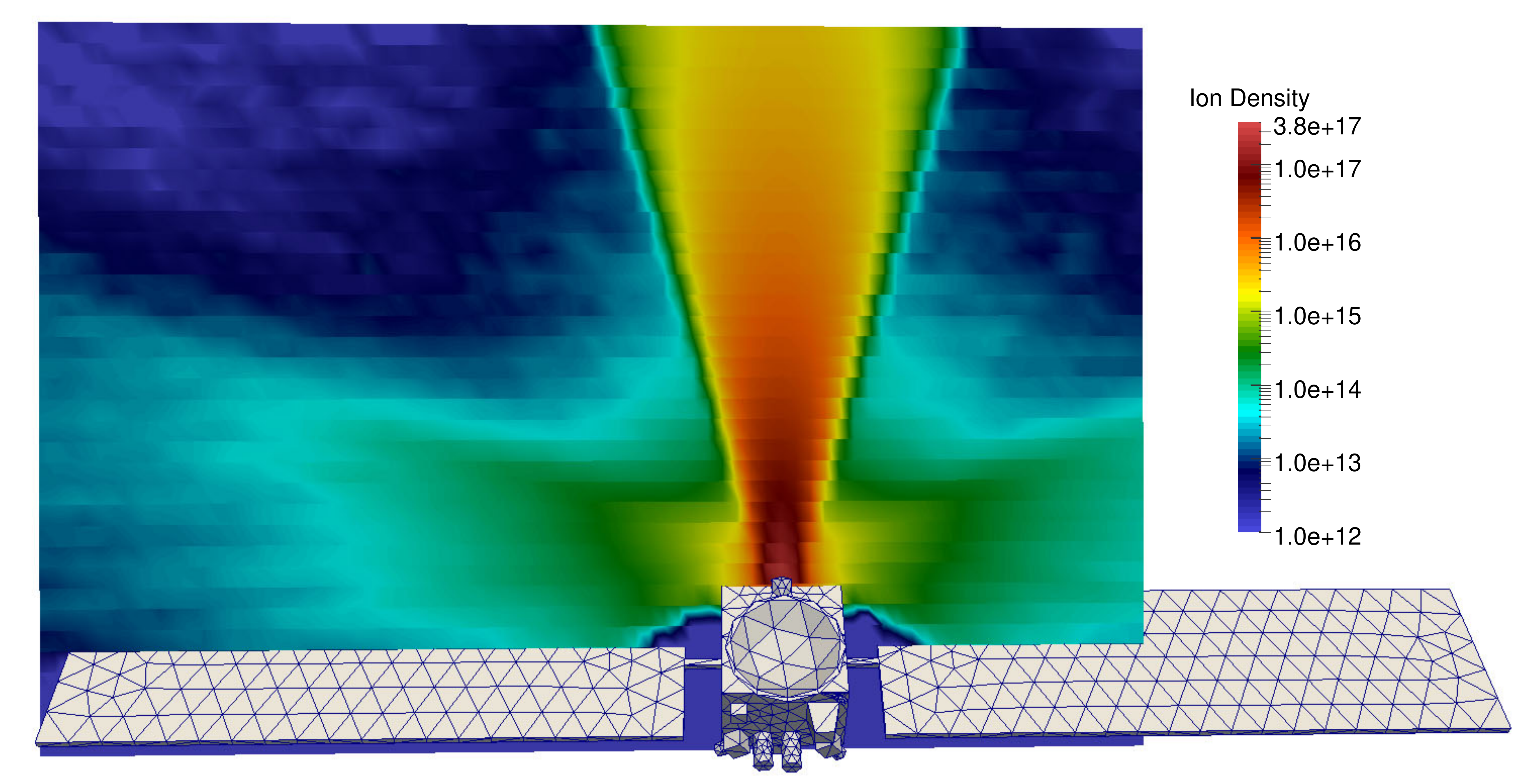}
    \caption{Ion density predicted with a hybrid plume model with polytropic electrons (inertialess, unmagnetized and collisionless) and $\gamma = 1.3$, in a complex geometry scenario, corresponding to the DAWN spacecraft. Charge exchange collisions for ions are included. Courtesy from S.J. Araki \cite{arak19}, IEEE Transactions on Plasma Science \textbf{47}, 4898–4908 (2019).}
    \label{fig: SC interaction complex circuit}
\end{figure}
In all cases, the potential and the currents flowing through these circuit nodes are solved through Kirkhoff's laws \cite{cich17,arak19}. Two main types of surface elements are generally considered in the model: (i) dielectric surfaces which, depending on the model, either force an equal ion/electron current from the plasma \cite{cich17} or are charged up by the net plasma current without transferring this accumulated charge elsewhere \cite{arak19}, and (ii) conductive/partially conductive surfaces which can exchange electric currents with other elements and are eventually connected to the spacecraft bus. When transients are requested and the electrical capacitance of the metallic surfaces is well known, implicit schemes to solve Kirkhoff's laws are employed to improve the stability of the modeled circuit, using time steps for the charge propagation that can be lower than the time step considered by the hybrid plasma plume model. Finally, while the electric currents to the surfaces are straightforward to compute from a full-PIC plume model, these are not when using a hybrid model. In this case, the electron contribution to the collected current is computed by either assuming the electron thermal flux $j_\mathrm{e0}$ (when the plasma potential $\phi$ is lower or equal to the surface potential $\phi_\mathrm{w}$) or by calculating the electron net flux crossing a thin plasma sheath (when quasi-neutrality holds and the sheath is not resolved in the model), i.e.
\begin{equation}
j_\mathrm{w,e} = j_\mathrm{e0} \exp{\left( -\frac{e \left(\phi-\phi_\mathrm{w} \right)}{T_\mathrm{e}} \right)}.
\end{equation}

\subsection{\label{subsec: plume-EMC} Plume-electromagnetic compatibility}

The impact of electrical thrusters on the telecommunications system of the satellite is a critical point when thrusters are firing. To be electromagnetically compatible, the radiations emitted by the propulsion sub-system must be under a certain level to avoid interference with the satellite. The emission of the radiated electric field has been measured on a wide range of frequency, from 10's of MHz to 10's of GHz frequencies corresponding to the antenna emission band, using semi-anechoic chambers (to prevent the thruster from the electromagnetic environment) juxtaposed to standard vacuum chambers where the thrusters operate\cite{Biet09, Holst20}. Specific electrical sensors that can be directly mounted in the vacuum facility simplifying the measurement system have also been recently proposed in the context of the HT\cite{Maz22HT} and VAT\cite{Maz22VAT}.

\begin{figure}[!ht]
    \centering
    \includegraphics[width=0.46\textwidth]{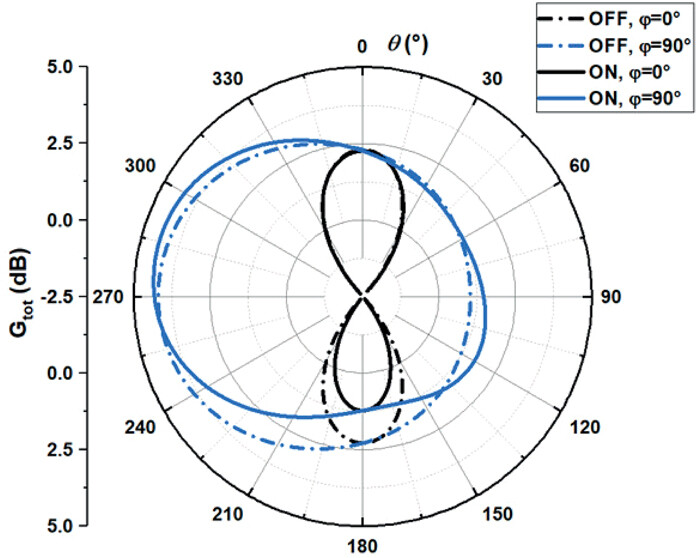}
    \caption{Antenna gain radiation pattern for two scenarios, in the antenna emission plane ($\varphi$ = 90 deg.) and in the plane perpendicular to the emission ($\varphi$ = 0 deg.) with (ON) and without (OFF) the plasma. Reproduced with permission from De Mejanes \textit{et al.}, Journal of Applied Physics \textbf{131}, 243303 (2022). Copyright 2022 AIP Publishing LLC.}
    \label{figIII.D}
\end{figure}

Furthermore, the charged particles that compose the thruster plume can alter the operation of the antenna by modifying the telecommunications or telemetry signal. This issue has been addressed in a recent modeling study in the context of a HT mounted on a small size satellite simplified mockup \cite{Meja22}. A 2D hybrid-PIC plume model called JET2D \cite{Garr02} has been adapted and validated against the ion angular current distribution measurements of a miniature HT \cite{Hall2020} to determine the profile of the electron plasma frequency (from the predicted density profile). Assuming a constant electron-neutral collision frequency, the profile of the electric permittivity (under the Drude's approximation \cite{Lieb2005} can be computed and used as input data for an electromagnetic solver (Ansys HFSS\cite{Ansys}). De Mejanes \textit{et al.} \cite{Meja22} have applied this versatile approach to a dipole antenna at 436 MHz on a 6U nanosatellite. For that specific configuration, no significant changes in the reflection coefficient nor the radiation efficiency were noticed when operating the thruster. Figure \ref{figIII.D} illustrates the weak deformation of the radiation diagram when the plasma is on.

\section{\label{sec: future challenges} Future challenges}
Here we present some interesting recent lines of research development aimed at improving the reliability and computational performance of EP particle-based simulations.

\subsection{\label{subsec: alternative propellants} Collisional database and alternative propellants}

Particle-based models allow an accurate description of electron and ion collisions with neutrals, a fundamental process to correctly describe the physics underlying and the global efficiency parameters of the different complex thruster configurations. Therefore, a predictive particle model not only requires a proper implementation of the numerical solution techniques, but also depends on the availability of reliable input data, such as electron- and ion-scattering cross-sections with neutrals.
Even if nowadays the number of experimental/theoretical groups measuring/calculating cross sections is rapidly diminishing, large data set (presented as look-up table with non-uniform energy intervals) are collected (often for total and rarely for differential cross sections) and available on different web platforms \cite{IAEA,lxcat,quantemol,NIST,NIFS}. For this reason, different approximations are used for the calculation of the anisotropic scattering angle valid for some electron-neutral \cite{Janssen16,Park22} and ion-neutral \cite{Wang17,Gueroult18} collisions in Monte Carlo simulations. 

\begin{figure}[!ht]
	\centering
	\includegraphics[width=0.47\textwidth]{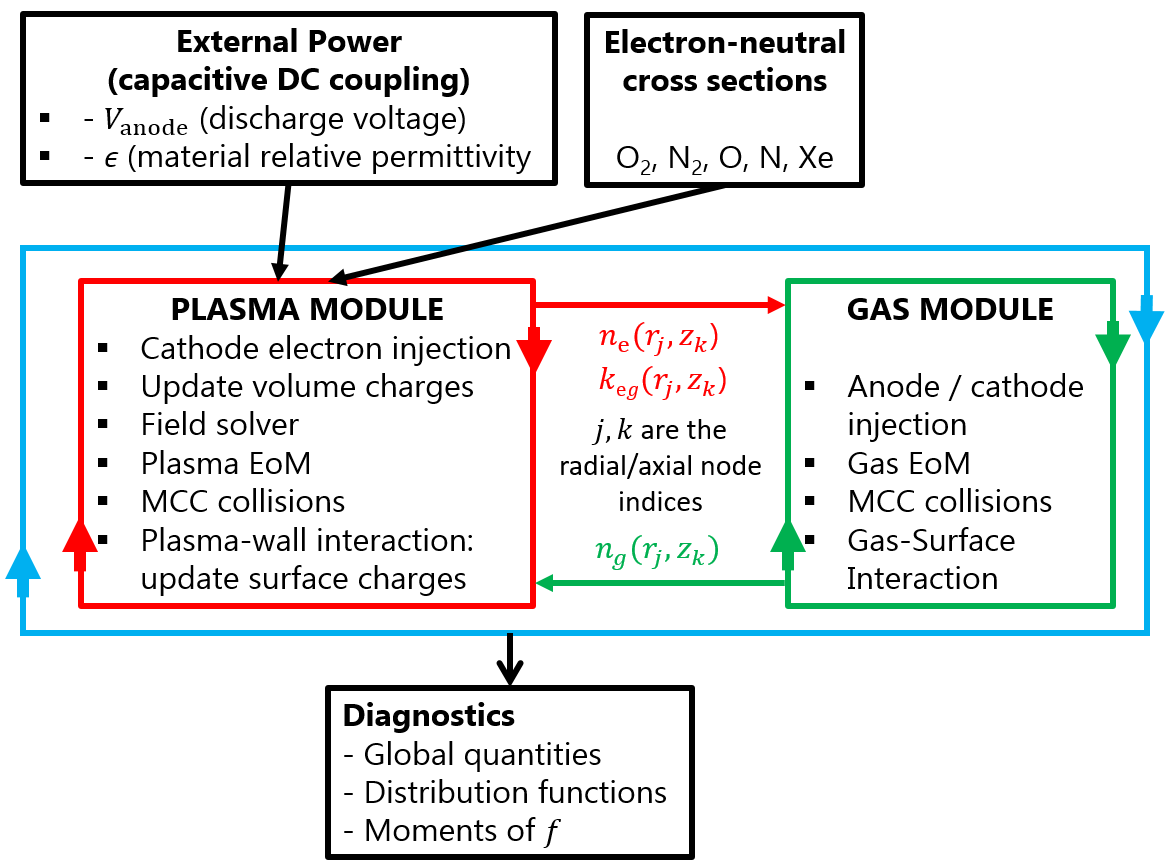}
        \caption{Scheme of a coupled PIC-MCC / TPMC simulation for a HT discharge simulation. The plasma (gas) modules see respectively the neutral gas (plasma) as fixed non-uniform backgrounds with density $n_g$ ($n_\mathrm{e}$), Reproduced with permission from Taccogna \textit{et al.}\cite{Taccogna2022}, Frontiers in Physics \textbf{10} (2022). Copyright 2022 Frontiers media SA.}
	\label{fig: PIC/TPMC loop}
\end{figure}

Simulation results are highly sensitive to the model used for neutral dynamics. It has been recently \cite{Pan23,Faraji23b} demonstrated that different numerical treatments of the neutrals change the spatio-temporal evolution of a HT discharge. Ion-wall recycling, neutral-wall reflection and energy accomodation coefficients and ion-neutral collisions cause a significant variation in the neutrals density and temperature maps.
Unfortunately, the plasma-neutral coupling is very difficult to implement due to the large disparity in their respective spatial and temporal characteristic scales. A first attempt \cite{Taccogna2022} has been made by developing a particle-based code with two different alternating modules for the plasma species (PIC/MCC) and for the neutral gas species (Test Particle Monte Carlo - TPMC), which are iterated and coupled until convergence. The two modules use their own grid and timestep to evolve their own Boltzmann's equation. The corresponding simulation loop scheme is reproduced in Fig.\,\ref{fig: PIC/TPMC loop}.

The plasma-gas coupling becomes fundamental to estimate thruster performances in case of molecular propellants. For air species ($\mathrm{N}_2-\mathrm{O}_2$ mixture relevant for the air-breathing concept \cite{Gurciullo19,Marchioni21,Taccogna2022,Munro23}), iodine $\mathrm{I}_2$ \cite{Lafleur2022,Andrews23}, water vapor $\mathrm{H}_2\mathrm{O}$ \cite{Sheppard2020,Shirasu23} and carbon dioxide $\mathrm{CO}_2$ \cite{Hohman12} (relevant for Venus and Mars atmospheres), the large variety of electron-molecule processes, the gas-wall interaction (tackled in Sec. \ref{subsec: plasma wall interaction}) and the molecular vibrational kinetics must be included for a realistic estimation of the ionization efficiency. In particular, the energy partition between atomic byproducts in the electron-induced molecular dissociation is very important for determining the possible impact of the subsequent ionization of hot atoms.

\begin{figure}[!ht]
	\centering
	\includegraphics[scale=0.49]{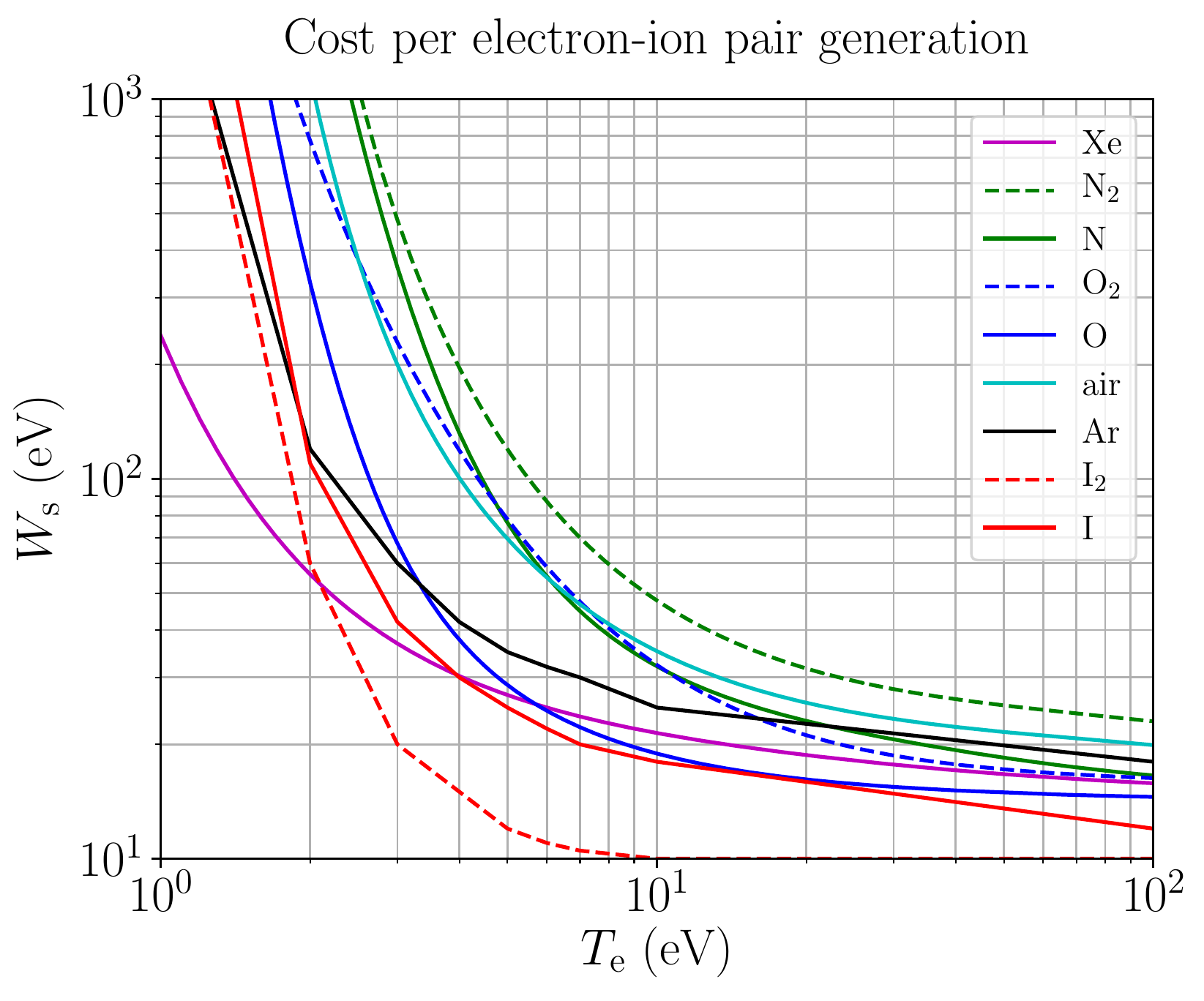}
        \caption{Collisional energy loss per electron–ion pair created, $W_\mathrm{s}$ (eV), versus the electron temperature $T_\mathrm{e}$ (eV) in Xe and air relevant species (N, N$_2$, O, O$_2$ and an air mixture 0.5 N$_2$/ 0.5 O) \cite{Taccogna2022} and in atomic (I) and molecular (I$_2$) iodine \cite{Lafleur2022}. Atomic species are shown by solid lines, while molecular species by dashed lines.}
	\label{fig:energycost}
\end{figure}

In this regard, in order to measure the ionization efficiency of a particular propellant, it is often useful to calculate the electron energy used for ionization compared with the energy dissipated in all inelastic processes. The global energy cost to create an electron-ion pair in the thruster discharge \cite{Garrigues2012} is defined as

\begin{equation}
W = \sum_{s=1}^{N_\mathrm{s}} \chi_s W_s
\end{equation}
where the index $s$ refers to the different gas species (atomic and molecular) of the propellant mixture, $\chi_s$ is the neutral fraction of the $s^\mathrm{th}$ neutral species and $W_s$ represents the corresponding cost to create an electron-ion pair defined as
\begin{equation}
W_s = \frac{\sum_{p=1}^{N_\mathrm{p}} \varepsilon_p k_p}{k_\mathrm{ion}}.
\end{equation}
Here, the numerator represents the total inelastic power losses, that is, the total energy loss due to all inelastic processes (including the ionization) involving the electron and the neutral specie $s$. $k_p$ and $\varepsilon_p$ are the inelastic rate coefficients and the energy thresholds for the inelastic process $p$, respectively. In the denominator, $k_\mathrm{ion}$ is the ionization rate coefficient of the process $e + \mathrm{S} \rightarrow 2e + \mathrm{S}^+$. Fig.\,\ref{fig:energycost} reports $W_s$ for different species: Ar, Xe, air species and iodine.

\begin{figure}
	\centering
	\includegraphics[scale=0.49]{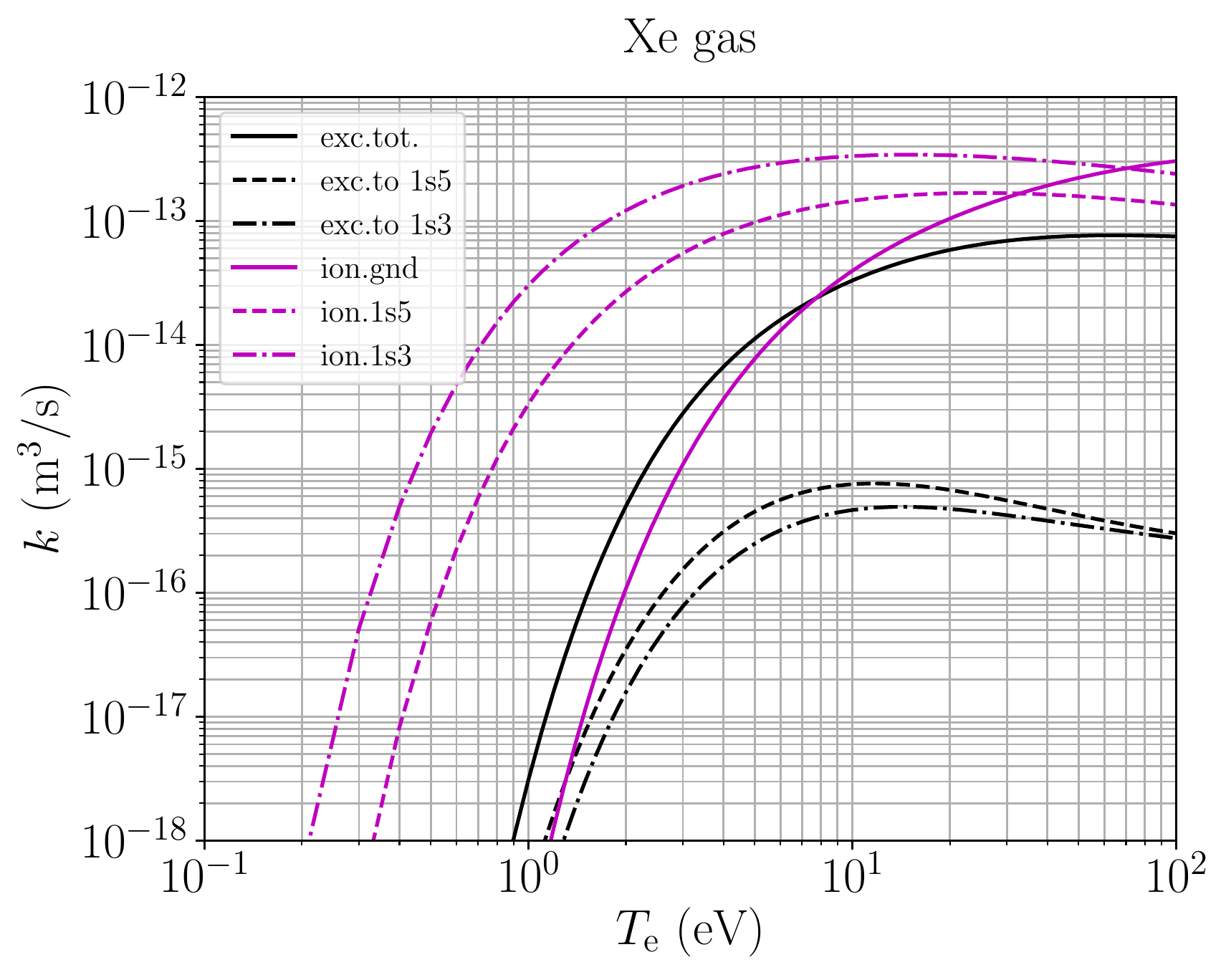}
        \caption{Xe excitation rate coefficients \cite{Hayashi03} from ground to any excited state (exc.tot.) and from ground to metastable states (exc.to 1s5, exc.to 1s3), and Xe ionization rate coefficients of the ground state \cite{Hayashi03} (ion.gnd) and of the two metastable states \cite{Hyman1979} (ion.1s5, ion.1s3). Paschen notation is here considered.}
	\label{fig: rates_Xe}
\end{figure}

Usually, all the atomic and molecular species are assumed to be in their electronic ground state since, in the low-pressure regime typical of electric thruster discharges, the radiative excited states decay to the ground state much faster ($\tau_{rad} \approx 10^{-9}$ s) than the electron collisional time. Therefore, the ionization process is always taken from the electronic ground state. This is not true when considering metastable states. Recent studies \cite{Zheng20,Wen21,Wen22,Yamashita2022} are starting to investigate the role of metastables (for Ar, Xe and O$_2$) and of vibrational kinetics (for molecular propellant) with particular emphasis on the impact of step-wise ionization. In this regards, often, state-selective (dependant on the initial vibrational state) and ionization of metastables cross sections are often missing in the literature. Fig.\,\ref{fig: rates_Xe} reports, for Xe, the excitation rate coefficients from ground to any excited state (exc. tot.) and to the existing two metastable states ($1s_5$ and $1s_3$, in Paschen notation). Additionally, the ionization rates of the ground state and of these two metastable states are also reported, showing that, at low electron temperatures, the contribution of metastable states to the total ionization rate can be relevant.
%

\subsection{\label{subsec: hpc techniques} HPC techniques}
The multiscale nature and large computational domains required in EP particle simulations have fostered the adaptation of existing codes to relatively new and highly efficient parallelization techniques \cite{Chaudhury19}. As an illustrative example, a 3D SPT-100 channel features a length of around 3 cm, and inner and outer channel radii of respectively 3.5 and 5 cm. For an average plasma density of $10^{18}$ m$^{-3}$ and an electron temperature in the order of 10 eV, the Debye length is approx. 20 $\mu$m, which means that a 3D simulation would require around 10 billion cells, and therefore hundreds of billions of particles (for a reasonable noise level). The corresponding memory to store information on these numbers of macro-particles is in the order of 10 TB of RAM memory, which clearly requires a distribution of the computational task over a large number of physically separated computational nodes.

\paragraph{Hybrid MPI-OpenMP parallelization} 
The most classical massive parallelization technique employs a particle and mesh decomposition through the combined ``Message Passing Interface'' (MPI) and ``Open Multi-Processing'' (OpenMP) parallelization techniques. 

The MPI parallelization technique has been developed in 1990s \cite{Dawson1993} and consists in running simultaneously the code on several computational nodes (or processes), which can exchange information with ``neighboring ones'' through communication over a network protocol. When using MPI, the simulation domain is generally split between these processes, and the number of neighbors of such a domain decomposition depends on the dimensionality of the problem: 2 neighbors in 1D, 8 neighbors in 2D, and 26 neighbors in 3D. A subdivision of the global domain between MPI processes permits to overcome the RAM memory issue introduced above, since each node now sees only a fraction of the global simulation data. Moreover, the maximum number of available cores for the computational task is no more limited to that of a single physical CPU (nowadays limited to 48-56 physical cores), so that an extremely large speed-up can be achieved (examples exist with speed-up in the order of 1000 \cite{ortw15}).

The drawback of using MPI is related to the overhead for network communication between processes, which depends on the amount of communicated data, on the possibility of performing non-blocking communication (which reduces the overhead) and on the latency and bandwidth of the network type. In particular, gigabit and InfiniBand are the most common network protocols, with the latter enabling larger transfer rates but being generally more expensive \cite{isma11}. Of course, particles are subdivided between MPI processes just like the simulated domain, and a dedicated communication function must be envisaged to exchange particle data whenever a particle leaves an MPI process sub-domain and enters that of a different process. In this respect, in order to reach the best performance, it is paramount to have a homogeneously distributed computational effort between the different processes, a fact that has led to the development of new MPI codes with dynamic load balancing \cite{arak22}.

MPI domain decomposition is normally coupled with OpenMP \cite{mine17}, a HPC technique for shared memory systems that is compatible with MPI and consists in multi-threading, i.e. in dividing the computational task of a single MPI process between several threads, whose number should not exceed the number of available physical cores for that specific process, unless hyper-threading is efficiently exploited. Many codes in literature for plasma thrusters and their plumes are actually parallelized with OpenMP alone, since its implementation is much easier than the MPI one. In the case of OpenMP, care must be put to avoid the existence of race conditions, or simultaneous access by different threads to the same RAM memory variable. These race conditions can produce bugs and memory errors, and are normally handled with several approaches, such as reduction operations (consisting in making thread-private copies of certain variables), or atomic/critical commands, that however significantly slow down the code execution.

Another very relevant HPC technique that should be pursued is the efficient use of the L2 cache memory. Modern CPUs are capable of performing very efficient operations and optimizing the access to the slow RAM memory. In order to take advantage of the CPU efficiency, operations such as charge deposition and field gathering should be vectorized, while for optmizing the RAM access, the computational data should be stored compactly. In particular, macro-particles that are close in physical coordinates (e.g. belonging to the same cell) should be stored closely also in the RAM memory (e.g. in the dedicated lists used in particles codes). A way to achieve this is through particle sorting algorithms \cite{garrigues2016}, which order the particles in terms of their occupied cells. 

Finally, especially in multi-scale simulations, in which the plasma density changes by several orders of magnitude across the domain, the overall computational cost can be reduced with ``Adaptive Mesh Refinement'' (AMR) techniques\cite{Jambunathan2018}, based on octrees.

\paragraph{Massive parallelization on GPU}
Graphics processing units (GPUs) represent very powerful hardware with a throughput-oriented architecture, with raw arithmetical performance and memory bandwidth by far exceeding those of the conventional central processing units (CPUs). However, GPUs can realize their potential only for a special class of problems, which can be partitioned into a large number of sub-problems on two levels. On the coarse level, the sub-problems should be independent of one another for the sake of scalability and are to be processed by blocks of threads, while, on the fine level, the sub-problems are concurrently processed by threads communicating between themselves via different means within the block. Due to the much larger number of resident threads compared to the number of physical execution units on GPUs and a very small overhead of switching between the threads, it is possible to hide the latency by quickly finding threads ready for execution and putting them on the physical units. The number of physical execution units on a single GPU is very large and can be more than a thousand. The reason is that GPUs do not need sophisticated control units predicting the execution path for each separate thread to reduce the overall latency as it is done on CPUs, relying instead on using a GPU analog of the ``single instruction multiple data'' parallel computing paradigm\cite{Hwu2022}. With GPUs having a shorter history of development compared to that of CPUs, GPU global memory controllers are not strongly bound by the requirements to be compatible with old protocols, which allows GPU engineers to experiment more with novel concepts resulting in significant boost of the global memory bandwidth, e.g., via the advanced HBM (high bandwidth memory) interface. In turn, the traffic to the global memory on GPUs can be decreased through the data reuse employing different programmable fast additional memory types if the PIC algorithm exhibits a sufficient degree of locality, reducing the need for a complicated and bulky system of automatically managed cache memory typically present on CPUs. The resources spared on the control units and large cache memories can be dedicated increasing the number of units efficiently used by GPUs.

The PIC algorithms are well parallelizable on GPUs, either in whole or just for the particle-related parts (particle pusher and the charge or current density assignment), which typically require most of the computational time. The latter option can also have certain advantages: due to the reduced size of the transferred data (in the charge and/or current density) and the relatively large CPU-GPU bandwidth on modern computers, the cumulative time penalty of calculating the fields on CPU is not high and can be hidden by overlapping with the execution of some other task (for example, the processing of MCC algorithms), which would keep CPU and GPU busy simultaneously. 

Parallelization of the particle-related parts of the PIC algorithm on GPUs is typically based on two different approaches. The first approach relies on particle sorting with respect to their positions (e.g., Ref.\,\onlinecite{mertmann_2011}), which enables to minimize the race conditions when calculating particle moments serving as field sources, binary collision-based models, and adaptive particle management. In addition, it strongly improves the data locality, which is beneficial for data reuse utilizing the automatic cache or manually controlled shared memory. The second approach spares the particle sorting but needs to use the atomic addition for the calculation of field sources (e.g., Ref.\,\onlinecite{juhasz_2021}). Due to the improved performance of atomic functions on modern GPUs, this can be a viable option if no additional algorithms demanding sorting are required.

An example of GPU parallelization of codes used for modeling propulsion-related problems can be found in Ref.\,\onlinecite{Gallo2021} for a single GPU and in Refs.~\onlinecite{Jambunathan2018,Jambunathan2018a,Nishii2022} for multiple GPUs. Besides that, one can find discussions of GPU parallelization techniques used in commercial codes\cite{Amyx2012,Leddy2018}. Popular tools for the GPU parallelization are the ``Compute Unified Device Architecture'' (CUDA) or ``Open Accelerators'' (OpenACC) (e.g., the latter has been recently used to parallelize the LTP-PIC code developed by the PPPL group and utilized previously for the benchmark in Ref.\,\onlinecite{Charoy_2019}). The massive GPU parallelization using CUDA was employed in the implicit energy-conserving ECCOPIC2S-M code used by the RUB team in the HT-related benchmark described in Ref.\,\onlinecite{Charoy_2019}. However, because of the benchmark requirements, the code was run using the same time and cell size as the other explicit PIC codes so that the benefits of the GPU parallelization became visible only in a different benchmark\cite{Villafana_2021}. There, also a GPU-parallelized but an explicit version of the same code used by the RUB team showed performance comparable to the best CPU-based code. Whereas it is hard to make direct performance comparisons, since the final performance depends on the number of GPUs and CPUs used, one can consider a heterogeneous CPU-based system enhanced with GPUs as a competitive cost-, size-, and energy-efficient alternative to large clusters based on CPUs only. Note also that GPUs offer numerous opportunities of utilizing additional computation units performing computations with a reduced precision, which can be combined with more accurate calculations in critical parts of an algorithm without noticeable overall precision losses. 

\subsection{\label{subsec: reducing cost} New numerical techniques for computational cost reduction}
In this section we present some ideas implemented in a classical PIC scheme to increase computational performances without losing the detail of the physics represented. In particular, the techniques used and verified in the modeling of electric thrusters will be shown.

\paragraph{\label{subsec: sparse-grid}
Sparse-grid method}
The sparse grid (SG) method is a specific discretization technique used for the interpolation of functions and resolution of partial differential equations \cite{Garc13}. The SG method is based on the construction of a system of sub-grids with a coarser resolution that will be used for the calculations. The solution of the problem can be reconstructed by the combination of the solution on each of the sub-grids with a minimum of error by using the so-called combination technique\cite{Ober21}. Recent studies have combined SG and ES PIC approaches with the combination technique. Ricketson \textit{et al.} \cite{Ricketson_2017} have demonstrated the proof-of-concept by studying the Landau damping in 2D and 3D, and the diocotron instability in 2D, assuming a collisionless plasma where ions remain at rest, and using purely periodic boundary conditions. Garrigues \textit{et al.} \cite{Garr21a, Garr21b} have extended the method to low temperature plasma discharges including collisions, charged particles motion, and Dirichlet boundary conditions.

In PIC models, the main source of error is the numerical noise or statistical error due to the particle sampling method\cite{BirdsallLangdon, Hockney_1988}. For the standard PIC approach, the total number of cells is $\mathrm{O}(N^{d})$, where ${N}$ is the number of grid points per direction and $d$ is the number of dimensions. The SG method employs $\mathrm{O}(N({\log N})^{d-1})$ grid cells, which is much lower than in the standard PIC algorithm. Using the same total number of particles leads to reduce the statistical error using SG techniques, but more interestingly, for the same statistical error, the total number of particles employed in the SG approach can be much less\cite{DelM2AN, MURA2021}. By diminishing the total number of particles, both the resources for memory storage and computational time are reduced (see Refs.\,\onlinecite{del22, del22GPU} for efficient parallelization of SG PIC algorithms on CPU and GPU shared memory architectures, respectively).

\begin{figure*}
	\centering
	\includegraphics[scale=0.3]{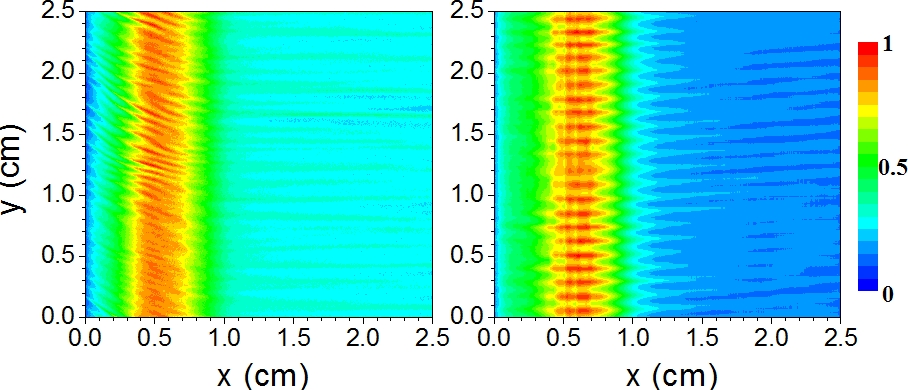}
        \caption{Axial \emph{x} - azimuthal \emph{y} simulations of the HT ECDI for a computational domain of 512 $\times$ 512 grid cells. 2D profiles of ion density at a given timestep, with (left) standard ES PIC, and (right) sparse grid ES PIC. The maximum ion density is $2.2\times{{10}^{17}}$ $\text{m}^{-3}$ and $2\times{{10}^{17}}$ $\text{m}^{-3}$, in left and right figures, respectively. The number of particles-per-cell is the same (400), and the total number of particles is ${{10}^{8}}$ and $5.2\times{{10}^{6}}$, respectively. Reproduced with permission from L. Garrigues \textit{et al.}\cite{Garr21b}, Journal of Applied Physics \textbf{129}, 153304 (2021). Copyright 2021 AIP Publishing LLC.}
	\label{figIV.C}
\end{figure*}

Fig.\,\ref{figIV.C} shows a comparison between the standard and SG ES PIC models in the context of the ECDI in a HT\cite{Garr21b}. The conditions are the same as in Fig.\,\ref{figII.A.2} except that the computational domain is now a square. The gain in computational time is 6.5 between standard and SG ES PIC models for identical computer resources. We see that the SG ES PIC algorithm is capable of reproducing the ion density profile and the ECDI instability but with a certain error that can be reduced using more sub-grids (but accompanied by a reduction of the computational gain). This is due to the numerical error associated to the mesh size of sub-grids and reconstruction using the combination technique that is not able to cancel out the non-mixture derivatives for non-smooth solution profiles. Works are underway to apply the over-sampled SG method detailed in Ref.\,\onlinecite{DelM2AN}, which should reduce this error, and to extend this method to 3D HT models \cite{Chung2022}.

\paragraph{\label{subsec: implicit PIC} Implicit PIC models}
The explicit PIC method (see Sec. \ref{sec: particle-based models}) evolves particle positions and velocities based on the information available from the previous time steps, which minimizes the number of calculations needed to make the update to the next time level. In particular, note that in the fully EM case, one calculates field components at the next discretized time instant from the corresponding time derivative, with all other quantities taken from the current one, which makes an algebraic matrix inversion unnecessary, provided the leapfrog algorithm is used.
The drawback of this approach is the corresponding decoupling of the particle and field evolution during the time step. If the latter exceeds a characteristic time scale intrinsic to the discretized field-plasma system, such as the plasma period or the time it takes an EM wave to cross a grid cell, the PIC algorithm becomes unstable due to the stiffness of the underlying differential equations. Another problem of the conventional explicit algorithm used in most cases is that the same function for the charge density assignment to the grid vertices and the reciprocal field interpolation leads to the ``finite-grid instability'' (FGI), causing numerical heating compromising or breaking down the simulation if the cell size becomes greater than the Debye length. For modeling discharges with large plasma densities or large dimensions, this can be a big problem since the required large computational grid will translate into a long computational time, especially for multi-dimensional simulations.

To circumvent these limitations, it was proposed to use implicit methods, which aim at the time integration algorithms evolving particles and fields in a coupled way. Although such methods typically require many more computations compared to the explicit PIC, their stability does not depend on having to resolve the Debye length, the time it takes for an EM wave to cross a grid cell, or the plasma period. As a result, the overall cost of an implicit PIC simulation modeling a dense plasma can be much smaller than that of an explicit PIC simulation. Two early implicit PIC approaches were proposed in the literature: the direct-implicit method (DIM) \cite{friedman_1981,Langdon1985} and the implicit moment method (IMM) \cite{mason_1981,Brackbill1985}. Although originally, these methods were formulated in a fully implicit form that requires an iterative procedure to converge, due to the low computational power of computers of that past period, they are most frequently used in a semi-implicit form requiring just a single iteration, as in the case of the explicit method.

The DIM linearizes equations describing the time evolution of the discretized field-plasma system with respect to the electric field at the new time instant, which allows expressing the charge and current densities as a sum of two parts, of which the first one is calculated using the previously known quantities and the second one is proportional to the new electric field. The latter can be combined with the other terms in the discretized Maxwell's equations, which contain the new electric field. The equations determining the time evolution are usually discretized in DIM using the leapfrog algorithm.

In contrast to this, the IMM estimates the new electric field from the momentum equation linearized with respect to the new electric field, albeit with the stress tensor calculated from the PIC data of the current time. The electric field estimate for the new time instant calculated in this way would differ from the new electric field obtained from Maxwell's equations, but if the time step is not very large, the discrepancy can be tolerated. The time integration scheme exploited in the IMM is based on the Crank-Nicolson algorithm. Even in the semi-implicit form, these algorithms allow breaking the curse of the time step limitation and using relatively large time steps. However, they suffer from the numerical heating/cooling related to the FGI and caused by the lack of energy conservation. Although this problem appears to be less acute compared to the explicit PIC and the implicit methods allow adjusting the amount of numerical error in the energy conservation (by varying the $\theta$ parameter used to mix the old and the new electric field values in the equations of motion and field equations), the FGI still plagues self-consistent simulations of large plasmas with high densities, where the cell size can become significantly larger than the Debye length.
Nevertheless, the DIM was used in the codes EDIPIC\cite{Sydorenko2006}, LSP\cite{Clark2005}, and Warp\cite{vay_2013} employed for simulations relevant 
to EP \cite{2014APS..DPPGP8110C,levko_2013,Kaganovich2019}, and in other codes (e.g., Ref.\,\onlinecite{lei_2022}). The IMM was used in Ref.\, \onlinecite{Cho2012} to model an SPT-type HT.

Recently, it was realized that if the IMM equations are implemented in their original form exactly (without the linearization), and if the functions used for the current density assignment and the electric field interpolation are the same, then for $\theta=0.5$, they conserve energy exactly, i.e., to a desired accuracy in a controllable way \cite{markidis_2011,chen_2011} (see also a discussion of the algorithm usage for the modeling of technological plasmas featuring electrodes and reactor walls, external networks, and collisions in Ref.\onlinecite{Eremin2022b}). Key ingredients for the energy conservation were the electric field update with the Ampère law even in the case of electrostatic simulations\cite{chen_2011,Eremin2022b} and employing the same shape function for the electric field and current density calculations.
It turned out that such energy-conserving PIC (ECPIC) algorithms have a different domain of parameters where the FGI is triggered compared to the momentum-conserving schemes \cite{Barnes2021}. Whereas the latter becomes unstable if the Debye length or the plasma period are not resolved by the cell size or the time step, the former suffers from the FGI only if the average electron drift exceeds their thermal velocity (note, however, that such a study was never conducted for magnetized plasmas). The FGI manifests itself differently in the energy- and momentum-conserving PIC codes. Due to the corresponding conservation properties, it can lead to the non-physical conversion of the momentum of a directed motion in the former\cite{langdon_1973} and the excessive numerical heating in the latter\cite{ueda_1994}. When  the FGI is not excited, the momentum (energy) is typically conserved to a tolerable accuracy in the energy-(momentum-) conserving PIC algorithm. Furthermore, such accuracy can be improved with subcycling in the orbit integration, which is also required to ensure the charge conservation in case the latter is needed \cite{chen_2011}. The absence of numerical heating and the FGI stability properties of the energy-conserving PIC implicit schemes enables them to be a very efficient tool in combination with non-uniform mapped grids \cite{chacon_2013}, or with unstructured finite elements \cite{Parodi2022}. If the energy-conserving schemes are not used for these cases, one needs to ensure that the FGI or CFL stability criteria are fulfilled, which might be difficult in a simulation with cell sizes sometimes varying over a few orders of magnitude. In this situation, when there are no strict limitations on the time step and the cell size caused by the algorithm stability considerations, the corresponding quantities should be chosen such so that the desired time and spatial scales of interest are resolved. It might seem problematic to use large time steps when the resolution of electron cyclotron rotation in the presence of a magnetic field is not needed, but the corresponding orbit integration algorithms have also been developed \cite{Ricketson2020}.

The fully implicit methods need an iterative procedure to make the field and the particle updates consistent over the time step. The system solved has the dimensionality defined by the field grid, since the current density can be calculated from the updated particle positions and velocities if the current iteration for the new electric field values at the grid is known. This is referred to as the ``particle enslavement''\cite{chen_2011,markidis_2011}. A modern option for the iterative procedure is the Jacobian-free Newton-Krylov method \cite{chen_2011}, which is economical in terms of the memory space and can be accelerated by using physics-based preconditioners\cite{Chen2014}. Alternatively, one can use the semi-implicit energy-conserving (ECSIM) algorithm recently proposed in Ref.\,\onlinecite{Lapenta2017} (see also Ref.\,\onlinecite{Lapenta2023}).
In modeling the plasma processing discharges, implicit energy-conserving PIC methods were already used to simulate ICP \cite{Mattei2017}, RF magnetron\cite{Eremin2023a,Eremin2023b,Berger2023}, and VHF CCP\cite{Eremin2022} plasmas and demonstrated good numerical properties. The ECCOPIC code employed in Refs.\,\onlinecite{Eremin2023a,Eremin2023b,Berger2023,Eremin2022} was verified in an international HT-related benchmark involving several codes\cite{Charoy_2019} and validated with experimental data\cite{Eremin2022,Berger2023}, demonstrating the method's functionality. 
Another work in this perspective direction indicating a plan to target specifically the EP applications, is presented in \onlinecite{Parodi2022}.

\paragraph{\label{subsec: reduced dimensional methods} Reduced dimensional order method}
A novel approach \cite{Faraji22,Faraji23} has been recently introduced for ES PIC based on the work done to model spokes in magnetrons \cite{Revel16}. It consists in approximating the 2D potential field $\phi(x,y)$ in terms of a superimposition of 1D potential fields along the $x$ and $y$ directions:
\begin{equation}
\phi(x,y) = \chi(x) + \eta(y).
\end{equation}
This allows substituting the solution of the 2D Poisson's equation with a system of two decoupled 1D Ordinary Differential Equations (ODE) for the functions $\chi(x)$ and $\eta(y)$. 
It effectively corresponds to solve in-parallel two 1D PIC simulations along the $x$ and $y$ coordinates with the simulations sharing the same macro-particles. Therefore, the total number of required macro-particles is essentially in the same order as in a 1D simulation and hence the computational cost, allowing to reduce the number of cells and the required total number of macro-particles from $\mathrm{O}(N^d)$ to $\mathrm{O}(dN)$, with $N$ being the number of cells along each direction, and $d$ the number of dimensions.

\begin{figure*}
	\centering
	\includegraphics[scale=2.5]{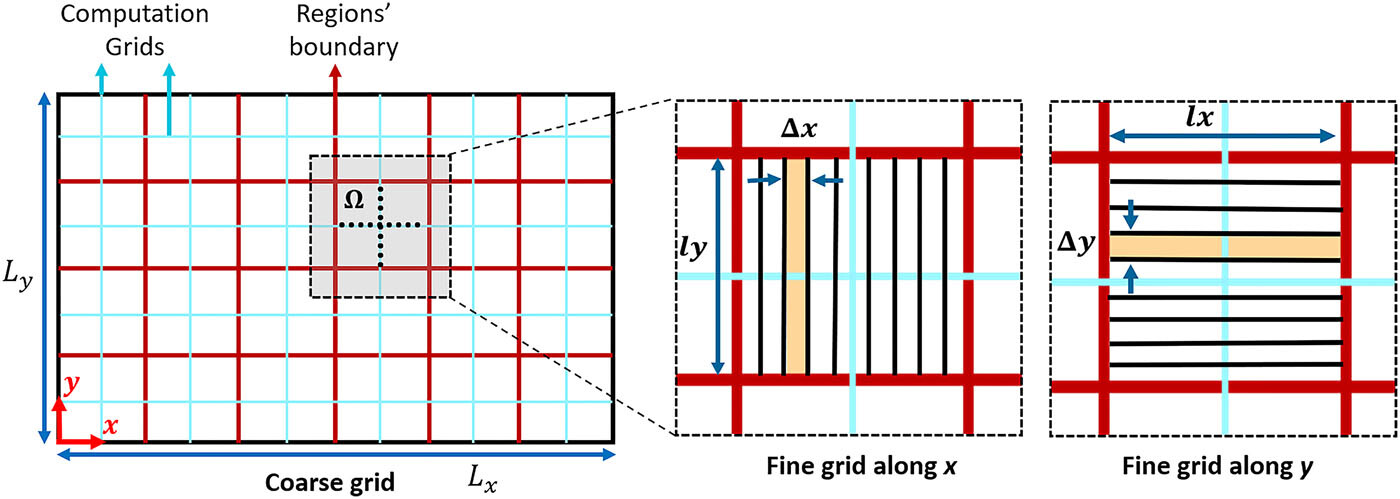}
        \caption{Schematic of the domain decomposition corresponding to the reduced-order PIC scheme: (left) decomposition into multiple “regions” using a coarse grid; (right) 1D computational cells for the discretization of each region along the $x$ and $y$ directions. The red lines indicate the boundaries of the regions, and the blue lines represent the computational grids along the $x$ and $y$ directions in each region. Reproduced with permission from Faraji \textit{et al.}\cite{Faraji23}, AIP Advances \textbf{13}, 025315 (2023). Copyright 2023 AIP Publishing LLC.}
	\label{fig: red_dim_PIC}
\end{figure*}

This is done by decomposing into multiple rectangular (represented in Fig.\,\ref{fig: red_dim_PIC} for a 2D case) or cubic (for 3D simulations) ``regions'', which can be thought of as the discretization of the domain using a “coarse grid”. Next, within any specific region, each simulation dimension is separately discretized using 1D “elongated” cells with the size criterion being the same as that in a conventional 1D PIC (i.e., smaller than the Debye length). This fine-discretization step yields a decoupling between the different coordinates in each region, allowing to reduce the number of cells and, hence, the required total number of macroparticles as anticipated above.

The accuracy and the computational cost of a reduced-order simulation depend on the fineness of the coarse grid, i.e., the number of regions to be used, which is problem-dependant. The pseudo-2D PIC scheme was shown to capture the multi-dimensional plasma phenomena in all 2D configurations relevant to a HT, i.e., axial-azimuthal, azimuthal-radial, and axial-radial \cite{Faraji22,Reza22,Faraji23,Reza23,Reza23b} with relatively few regions used, showing a computational speed-up with respect to the 2D simulation up to 50.
The method can be promising to reduce the full 3D picture into a 2D model making the full dimensional kinetic representation of HTs affordable. In fact, the expected speed-up for a quasi-3D simulation with respect to the corresponding classical full-3D one is about 2500.

\subsection{Deep Learning-Based Particle-in-Cell Method}
The next generation of PIC models can be integrated with Machine-Learning (ML) and Deep-Learning (DL) algorithms to directly calculate the electric and magnetic fields from the charged particle phase space, thus bypassing the Maxwell's equations solvers (Poisson's solver for the ES case).

In fact, ML and DL methods have emerged as valuable tools for data analysis and to replace or complement more traditional computational approaches. An example of such efforts is the development of DL-based pre-conditioners and linear solvers \cite{Fujita21}, heterogeneous linear solvers for Partial Differential Equations \cite{Markidis21}, use of DL-based methods in Computational Fluid Dynamics (CFD) solvers \cite{Guo16}, or weather forecasting \cite{Shao22}.

Recently \cite{Aguilar21}, a methodology to embed a DL electric field solver into an explicit PIC method has been developed and applied to the two-stream instability. The DL electric field solver is trained using the particle phase space information and its associated electric field.
The DL-based PIC method replaces then the interpolation step for the charge calculations and electric field solver of the traditional PIC method with two new steps: an interpolation of particle velocity and position into a phase space grid and a DL electric field solver that is the result of a DL neural network training. In this regard, several choices are possible: Multilayer Perceptron (MLP), Convolutional Neural Network (CNN) or Residual Network (ResNet), among others.
In order to train the DL electric field solver, a training data set, formed by the phase space grid and the associated electric field has been produced by running highly accurate traditional PIC simulations.

An application of the ML techniques in computational EP physics has also been focused on developing data-driven surrogate models \cite{Lee23} using the data to find closure models for plasma fluid system of equations. In particular, an ML algorithm \cite{Jorns2018} has been developed to fit the anomalous electron collisional frequency in HT to be used in the generalized Ohm's law for creating predictive hybrid models. The DL algorithm ultimately yields the following functional forms for the anomalous frequency,
\begin{equation}
\nu_{ano} \propto \omega_\mathrm{c,e} \bigg( \frac{u_\mathrm{i}}{v_\mathrm{d,e}} \bigg)^2,
\end{equation}
demonstrating the importance of the electron $\bm{E} \times \bm{B}$ drift $v_\mathrm{d,e}$ as the driving source for the onset of the turbulence and the ion drift speed $u_\mathrm{i}$ as its main saturation mechanism.

\section{\label{sec: conclusions} Conclusions}
This perspective paper has outlined the efforts in developing particle-based models of the different electric propulsion concepts since the first attempts in the early 90s up to the latest sophisticated multi-dimensional simulations.
In particular, the different electrostatic and electromagnetic schemes of the PIC/ Monte Carlo method have been presented in relation to the two different families of electric thrusters.
The modular nature of this method has allowed to model the vast variety of mechanisms contributing to the realistic representation of an electric thruster; the interaction between the electromagnetic / electrostatic fields and the charged particles leading to the electron heating and ion acceleration, the plasma-gas coupling, which is crucial in low pressure devices characterized by a high degree of ionization, and, finally, the different surface processes representing not only a particle energy loss term but also a source/recycling term. These mechanisms have all been discussed and the different dedicated algorithms presented.
Moreover, an outline of the current particle simulations methodology for plasma plume expansion and interaction with the satellite has been provided covering different topics such as hybrid versus full-PIC simulation approaches, circuit models for plasma-satellite interaction, and methods to assess the plume-electromagnetic compatibility.

An overview of the challenges, which according to the authors, will help push forward the boundaries of particle simulations in electric propulsion, has been presented. One of them is the availability of a collisional database for complex plasma chemistry (alternative propellants, more complete representation of vibrational and rotational kinetics, etc.) and the associated issue of a self-consistent simulation of the neutral gas phase in a affordable manner (given the extremely different time scales characterizing the plasma and gas phases). The constantly increasing cost of high fidelity simulations then urges the adaptation of all codes to HPC techniques, such as the hybrid OpenMP-MPI parallelization for CPU-based architectures, and the massive GPU parallelization, possibly combined with MPI to further boost the performance. Apart from HPC techniques, the feasibility of simulations will also depend on new methods to reduce their computational cost, such as the sparse-grid method, the novel implicit/semi-implicit energy-conserving PIC models, or the reduced dimensional order methods. Finally, the possibility of integrating future PIC simulators with Machine-learning algorithms has been thoroughly discussed.

The thrusters classification considered in this perspective paper has been based on modeling needs, thus differentiating PIC models requiring to solve only Poisson's equation (ES models) from those requiring to include a larger set of Maxwell's equations (EM models).

Regarding ES electro-spray thrusters, a major point to be addressed in future work is the droplet formation process, which can enable more precise injection conditions for PIC models of the extraction and acceleration process. Electrostatic gridded ion engines, on the other hand, still lack a comprehensive simulation of both the internal discharge chamber and the ion beamlet formation/coalescence process, possibly combining a full-PIC simulation of the former and a hybrid simulation of the latter. In $\bm{E}\times \bm{B}$ thrusters, the main open point is that of the fundamental characterization of the anomalous transport, through instabilities such as the electron cyclotron drift, modified two-stream, ion transit-time, or gradient drift (modified Simon-Hoh, lower-hybrid, or ion-sound) instability, sometimes also leading to the emergence of large-scale self-organized structures such as spokes, significantly enhancing the electron cross-field transport, or complex near-wall conductivity effects, coupled with secondary electron emission processes.

For what concerns EM thrusters, a full simulation retaining all terms in the coupled Maxwell-Newton system being evolved in the time domain and modeling high-density plasmas with realistic sizes remains elusive, although it is rapidly become feasible with modern high performance super-computing clusters featuring heterogeneous CPU/GPU nodes and algorithms tolerating large time steps (implicit and semi-Lagrangian) and the Debye length (implicit). Above all, it would allow to secure a self-consistent description of the complicated power absorption mechanisms, which are essential to the entire discharge physics. Such mechanisms may involve nonlinear particle-wave interaction, mode conversion, parametric and other instabilities, which are yet to be investigated in the context of the physics of thrusters. 
Although other aspects of the EM thrusters, such as the physics of plasma expansion in magnetic nozzle, are less dependent on the fully electromagnetic treatment, an appropriate evaluation of the power absorption and ionization profiles is substantial for making reliable quantitative assessments of the overall thruster efficiency.

Coming now to the topic of plasma plume expansion and interaction with the spacecraft, full-PIC simulations have started to emerge and highlight the deficiencies and limitations of hybrid particle models, most of the time based on the quasi-neutral assumption coupled with polytropic electron thermodynamics. This approach is clearly incapable of capturing both the complex electron cooling phenomena and the peripheral plasma behavior, which can strongly affect the production of slow charge-exchange ions constituting most of the ion backflow towards the satellite. Nevertheless, a complete full-PIC simulation coupled with a complex circuit model of the satellite is still a prerogative of hybrid models, and will require a great improvement of the existing code parallelization. In this context, hybrid models solving for the electron energy balance equation might prove to be an important alternative to full-PIC ones, although only the latter retain all the relevant kinetic effects.

Omnipresent in all PIC models of both electric thrusters and their plasma plumes is then the topic of the gas-wall and plasma-wall interaction. The main open issue here is the lack of accurate data at low impact energies for both the secondary electron emission yield and the ion sputtering yield, as well as the energy and angle distributions of the emitted particles (e.g. backscattered/true secondary electrons and sputtered atoms). Besides, it is also essential to characterize with dedicated experiments both the neutral reflection and ion recombination energy accommodation coefficients, which have demonstrated to have a non-negligible influence on the propellant utilization efficiency of certain thrusters, and, more in general, on the distribution of plasma and gas bulk properties inside.

Continuing on the line of elementary processes data, it is necessary to acquire more detailed data, such as differential cross sections (for the scattering angle in elastic collisions and for the energy of secondary electrons in ionization events), vibrational state selective (for molecular propellants) and ionization state selective (e.g. for metastable states) cross sections. The existence of metastable states, in fact, might have a non-negligible influence on the overall ionization process, a fact that is seldom checked in the vast majority of present-day codes.

In order to tackle the problem of the ever-growing computational cost of electric propulsion simulations, HPC techniques have begun to emerge that take advantage of heterogeneous multi-core CPU/GPU hardware on computing nodes of most modern HPC clusters. On the node level, one can combine both types of hardware by identifying parts of the algorithm which have a lot of branching points, hard-to-resolve data dependencies, and/or an inherently non-local global data access pattern. These parts would fit better for CPUs, whereas algorithmic parts with a large number of sub-problems requiring a similar instruction and data access pattern to solve, are better off placed for execution on GPUs. On a coarser level, the problem can be upscaled by harnessing multiple CPU threads or multiple GPUs on a single node and/or multiple nodes using the MPI set of instructions. The fast development of GPU hardware and software, less affected by compatibility issues compared to the CPU analogs, offers a lot of new perspectives concerning code parallelization. The development of novel algorithms benefiting from the large intra- and inter-GPU memory bandwidth and employing various reduced-precision units in mixed-precision algorithms is an interesting direction to go.

Additionally, one can cut computational costs by employing novel numerical techniques. One of the corresponding options is to use algorithms tolerating large time steps and/or large cell sizes. In this respect, the energy-conserving implicit or semi-implicit methods appear to be particularly promising, especially for multi-dimensional simulations. They have already proven their feasibility in the first electrostatic and electromagnetic PIC studies of large/dense technological plasmas, but there are much fewer reports of their use for thruster-related simulations. Another interesting technique is the Sparse PIC approach, which offers unquestionable advantages in terms of footprint memory, thanks to the smaller number of macro-particles for the same statistical error as in standard PIC methods, as well as a reduced size of mesh-related arrays.
If the strategy optimizations have been proposed and validated for shared memory architectures (CPU and GPU), work remains to be done for an optimized coupling between shared and distributed memory machines, especially for 3D algorithms. Furthermore, the transition from the Sparse PIC approach in its explicit version to the implicit Sparse PIC algorithms would offer an incredible gain in computational time. However, the conservativeness of the algorithm properties (total momentum, total charge, etc.) must be proven first. Finally, very recent works have demonstrated the capabilities of a new technique, known as ``reduced dimensional order'' method, which decouples the local field solution along the coordinate directions, and could yield total speed-ups in the order of $10^3$ in the most time consuming 3D scenarios.

To conclude, the combination of plasma propulsion and particle-based models has already proven to be very effective, representing an excellent example in which computational physics can fully express its potential. Thus, we expect that, in the next few years, PIC/Monte Carlo models will open up new scenarios in understanding physics-based questions (such as the existence of self-organized structures in the plasma) related to electric propulsion and, as a consequence, in the development of more efficient thrusters, with a particular attention to the study of power scaling up and down solutions and to the use of alternative propellants (from in-situ resources or condensable propellants). 
Our opinion is that the wide variety of available algorithms and models in the family of PIC/Monte Carlo techniques, the availability of new and more detailed data on volume and surface elementary processes, and the growing performance of computational platforms will together pave the way for the full numerical optimization and design of plasma thrusters.


\section*{Acknowledgments}
This material was based upon work supported by the Italian Ministry of University and Research (MUR) under the project PON ``CLOSE to the Earth'', No. ARS01–00141; F.T. and F.C. acknowledge fruitful discussions with P. Minelli, D. Bruno, V. Laporta and J. Zhou.
D.E. appreciates fruitful discussions with P. Parodi and R. Rakhimov.
G.F and L.G acknowledge the access to the HPC resources of the CALMIP supercomputing center under the allocation 2013-P1125; G.F and L.G are very grateful to their colleagues J.P. Boeuf, G. Hagelaar, and S. Tsikata for intense and fruitful discussions.

\section*{Data Availability Statement}
The data that support the findings of this study are available from the corresponding author upon reasonable request.

\section*{Authors contributions}
All authors have contributed to both original draft writing and review of this perspectives paper.

\appendix

\bibliography{papers}

\end{document}